\documentclass[lettersize,journal]{IEEEtran}
\usepackage{amsmath,amsfonts}
\usepackage{algorithm}
\usepackage{algpseudocode} 
\usepackage{array}
\usepackage[caption=false,font=normalsize,labelfont=sf,textfont=sf]{subfig}
\usepackage{textcomp}
\usepackage{stfloats}
\usepackage{booktabs}
\usepackage{url}
\usepackage{verbatim}
\usepackage{graphicx}
\usepackage{amssymb}
\usepackage{cite}

\usepackage{booktabs}
\usepackage{pifont,xcolor} 
\newcommand{\cmark}{\textcolor{black}{\ding{51}}} 
\newcommand{\xmark}{\textcolor{gray}{\ding{55}}}  
\usepackage{makecell} 

\begin{document}

\title{PAPR-Aware Waveform Design for Energy-Efficient MIMO-OFDM SWIPT}

\author{Chongda Huang,~Yue Xiao,~\IEEEmembership{Member,~IEEE,} 
        Qianzhen Zhang,
        Lilin Dan,~\IEEEmembership{Member,~IEEE,} 
        Xianfu Lei,~\IEEEmembership{Member,~IEEE,} 
        and Kai-Kit Wong,~\IEEEmembership{Fellow,~IEEE}%
\thanks{C. Huang, Y. Xiao, Q. Zhang, and L. Dan are with the National Key Laboratory of Wireless Communications, University of Electronic Science and Technology of China, Chengdu 611731, China (e-mail: chongdahuang@std.uestc.edu.cn, qianzhenzhang@std.uestc.edu.cn, xiaoyue@uestc.edu.cn, lilindan@uestc.edu.cn).}%
\thanks{X. Lei is with the School of Information Science and Technology, Southwest Jiaotong University, Chengdu 610031, China (e-mail: xflei@swjtu.edu.cn).}%
\thanks{K.-K. Wong is with the Department of Electronic and Electrical Engineering, University College London, London WC1E 7JE, U.K., and also with the Department of Electronic Engineering, Kyung Hee University, Yongin-si, Gyeonggi-do 17104, Republic of Korea (e-mail: kai-kit.wong@ucl.ac.uk).}%
}

\markboth{Journal of \LaTeX\ Class Files,~Vol.~14, No.~8, August~2021}%
{Shell \MakeLowercase{\textit{et al.}}: A Sample Article Using IEEEtran.cls for IEEE Journals}


\maketitle

\begin{abstract}
Simultaneous wireless information and power transfer (SWIPT) critically depends on waveform design, which governs both reliable data delivery and efficient energy harvesting. Among waveform characteristics, the peak-to-average power ratio (PAPR) plays a pivotal role: low-PAPR signals improve power amplifier (PA) efficiency, while high-PAPR signals exploit rectifier nonlinearities to boost harvested energy. This duality makes PAPR a fundamental design challenge in SWIPT systems. To tackle this issue, we establish a unified analytical framework that characterizes the PAPR-dependent behaviors of both the PA and the rectifier, thereby revealing how waveform statistics determine end-to-end energy transfer efficiency. Building on this insight, we propose a frequency-domain resource allocation strategy for power-splitting SWIPT, where spectral segments are adaptively assigned to balance communication throughput with energy harvesting performance. Here, a key contribution is to extend SWIPT to MIMO-OFDM architectures. Despite concerns over excessive PAPR in large-scale antenna–subcarrier configurations, we demonstrate that appropriate waveform adaptation and resource optimization can transform MIMO-OFDM into an energy-efficient platform for joint data and power transfer. Finally, simulation results confirm significant improvements in PA efficiency, rectifier output, and overall energy transfer, thereby validating the practical benefits of the proposed approach.
\end{abstract}


\section{Introduction}
\IEEEPARstart{S}{imultaneous} wireless information and power transfer (SWIPT) has emerged as a promising paradigm to address the dual requirements of reliable data communication and sustainable energy supply~\cite{choi2020simultaneous,krikidis2014swipt,perera2018survey,hossain2019survey} for next-generation wireless systems. By enabling devices to harvest energy and decode information from the same radio-frequency (RF) signal, SWIPT offers a practical solution to the energy constraints of battery-powered nodes such as internet of things (IoT) devices, wireless sensor networks, and wearable electronics. Its feasibility has been demonstrated in radio-frequency identification (RFID) systems, where passive tags scavenge energy from the interrogator’s signal to sustain operations and enable backscatter communication. Moreover, ongoing research has further explored advanced optimization and resource allocation strategies for SWIPT~\cite{8636993,6774838}. As wireless networks evolve toward ultra-dense, sustainable, and autonomous infrastructures, the seamless integration of information transfer and wireless power delivery is anticipated to play a pivotal role in their development~\cite{clerckx2018fundamentals,clerckx2018wireless}.

Over the past decade, extensive research has established the foundation of SWIPT by exploring fundamental rate-energy trade-offs, system architectures, and experimental validations~\cite{zhang2013mimo,kim2016new,jang2020novel}. For example, the seminal work in~\cite{zhang2013mimo} demonstrated the feasibility of multiple-input multiple-output (MIMO) broadcasting for SWIPT, which was later advanced through innovations in receiver design~\cite{choi2020simultaneous}, frequency-splitting techniques~\cite{jang2020novel},  peak-to-average power ratio (PAPR)-aware transmission strategies~\cite{kim2016new}, and multiuser beamforming schemes~\cite{xu2014multiuser}. More recently, waveform design has emerged as a central enabler for practical SWIPT, as highlighted in surveys and tutorials emphasizing its critical role in capturing nonlinear energy-harvesting characteristics~\cite{clerckx2018fundamentals,clerckx2018wireless}. In parallel, waveform design for wireless power transfer (WPT) has focused on multisine excitation and adaptive waveform synthesis to enhance rectifier efficiency~\cite{valenta2015theoretical,eidaks2022fast,clerckx2016waveform,clerckx2018fundamentals}, supported by the advances in rectenna modeling and optimization frameworks that accurately characterize nonlinear device behavior~\cite{abeywickrama2021refined,zhang2023waveform}. These transmitter- and receiver-side perspectives collectively point toward the need for a unified treatment, where the dual role of PAPR on both PA efficiency and rectifier performance can be jointly exploited to improve end-to-end system efficiency.

Among the available waveform candidates, orthogonal frequency-division multiplexing (OFDM) has become the cornerstone of modern wireless communications owing to its high spectral efficiency, robustness against multipath fading, and flexible resource allocation~\cite{andrews2014will,parkvall2018nr}. However, its adoption in SWIPT remains limited. The main challenge arises from the inherently high PAPR of OFDM signals~\cite{han2005overview,jiang2008overview}, which significantly impacts the energy efficiency and hardware design of integrated SWIPT transceivers~\cite{kim2016new}. On the transmitter side, high PAPR necessitates a large output back-off (OBO) in the power amplifier (PA) to maintain linearity, thereby reducing PA efficiency and increasing power consumption~\cite{raab2002power,camarchia2015doherty,zhang2023waveform,clerckx2018wireless}. On the receiver side, however, large waveform peaks enhance the nonlinear switching behavior of diode-based rectifiers, thus improving energy harvesting efficiency~\cite{clerckx2016waveform,collado2014optimal,valenta2015theoretical,ayir2023}. This dual yet conflicting role of PAPR creates a fundamental design dilemma in SWIPT: while low PAPR favors PA efficiency and communication reliability, high PAPR facilitates rectification and boosts harvested energy~\cite{litvinenko2018usage,ayir2021joint}.

This trade-off becomes even more pronounced in multi-antenna systems. In particular, MIMO-OFDM architectures—widely regarded as the backbone of next-generation wireless networks—tend to exhibit exacerbated PAPR due to the superposition of multiple transmit streams and the fine-grained resolution across subcarriers. As a result, many existing SWIPT designs deliberately avoid OFDM to circumvent these difficulties, albeit at the cost of reduced spectral flexibility and incompatibility with established communication standards. While significant efforts have been devoted to mitigating PAPR in OFDM for reliable communication—through approaches such as tone reservation~\cite{li2011improved,jiang2015novel,wang2019scr,li2018tone}, active-set optimization~\cite{krongold2004active}, selective mapping, and coding~\cite{claessens2019multitone}—the potential benefit of high-PAPR MIMO-OFDM waveforms for wireless power transfer has received far less attention. To date, only a few studies, most notably the multisine-based waveform design for WPT in~\cite{clerckx2016waveform}, have explicitly leveraged high PAPR to improve rectifier efficiency. This highlights a critical research gap: despite its ubiquity in communication systems, MIMO-OFDM remains underexplored as a candidate waveform for integrated information and power transfer, particularly from the perspective of enhancing energy delivery at the receiver.

\begin{figure}[!t]
\centering
\includegraphics[width=8cm]{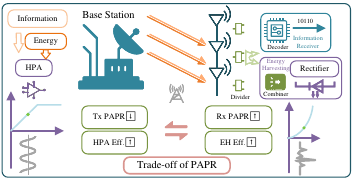}
\caption{Transmit–Receive Trade-off of PAPR in SWIPT Systems.}
\label{papr_aware}
\end{figure}

\begin{table}[!t]
\centering
\caption{Comparison of Related Works on PAPR-Aware SWIPT}
\label{tab:literature_comparison}
\begin{tabular}{lccc}
\toprule
\textbf{Reference} & \makecell{\textbf{Tx: PAPR}\\\textbf{\& PA Eff.}} & \makecell{\textbf{Rx: PAPR}\\\textbf{\& Rectifier Eff.}} & \makecell{\textbf{Applicability to}\\\textbf{Large-subcarrier}\\\textbf{OFDM}} \\
\midrule
\cite{jang2020novel,krikidis2020information} & \cmark & \xmark & \xmark \\
\cite{litvinenko2018usage,ayir2021joint,collado2014optimal} & \xmark & \cmark & \xmark \\
\cite{clerckx2018fundamentals,abeywickrama2021refined,zhang2023waveform} & \cmark & \cmark & \xmark \\
\cite{ng2013wireless,zhou2014wireless} & \xmark & \xmark & \cmark \\
\midrule
\textbf{This work} & \cmark & \cmark & \cmark \\
\bottomrule
\end{tabular}
\end{table}

Table~\ref{tab:literature_comparison} summarizes representative studies on PAPR-aware SWIPT. Prior works have typically addressed either transmitter-side PA efficiency or receiver-side rectifier performance, while only a few considered both simultaneously. Moreover, most of these studies are not directly applicable to large-scale OFDM systems. Fig.~\ref{papr_aware} further illustrates the dual yet conflicting role of PAPR in SWIPT, highlighting the need for a holistic design approach. In contrast, this work jointly accounts for both transmitter and receiver nonlinearities and proposes a framework scalable to MIMO-OFDM.

Against this background, the contributions of this paper are threefold:  

\begin{itemize}
  \item \textbf{Unified modeling framework:} A unified framework is established that explicitly characterizes the nonlinear, PAPR-dependent behaviors of both the PA and the rectifier, thereby providing analytical insights into the fundamental energy–information trade-offs of SWIPT systems~\cite{ayir2021joint,litvinenko2018usage,ayir2023}.  

  \item \textbf{Flexible resource allocation:} Building on this framework, a SWIPT architecture is proposed that integrates a power-splitting receiver with a frequency-domain resource allocation strategy, where distinct spectral segments are tailored to improve PA efficiency, enhance energy harvesting, and sustain data throughput~\cite{jang2020novel,krikidis2019tone,ng2013ofdm}.  

  \item \textbf{Energy-oriented metric and validation:} An energy-oriented signal metric is introduced to link waveform features to end-to-end energy delivery, enabling system-level evaluation and optimization. The proposed framework is validated through extensive simulations and experimental evaluations, showing consistent gains in amplifier efficiency, rectification performance, and overall power delivery compared with conventional OFDM-based SWIPT schemes~\cite{kim2020signal,eidaks2022fast}.  
\end{itemize}

The remainder of this paper is organized as follows. Section~II develops mathematical models for the PA and rectifier, explicitly relating their energy efficiencies to signal PAPR and introducing the proposed energy-oriented metric. Section~III presents the SWIPT system architecture and the frequency-domain subcarrier allocation strategy. Section~IV reports simulation results that validate the theoretical analysis and highlight the performance gains of the proposed design. Finally, Section~V concludes the paper.

\section{Modeling of Wireless Power Transfer}

\subsection{Power Amplifier Modeling}

The PA is the primary determinant of the transmitter's energy efficiency. To maintain analytical rigor and clarify the power conversion process, we define $P_{\text{DC}}$ as the consumed DC supply power and $P_{\text{out}}$ as the average RF output power. The drain efficiency is thus formally expressed as $\eta_{\text{PA}} = P_{\text{out}} / P_{\text{DC}}$. 

In practical LDMOS-based PAs, the operating point is characterized by the output back-off (OBO), defined as the ratio of the maximum saturated output power $P_{\text{sat}}$ to the average output power $P_{\text{out}}$:
\begin{equation}
\text{OBO} = 10 \log_{10} \left( \frac{P_{\text{sat}}}{P_{\text{out}}} \right) \text{ dB}.
\end{equation}

For MIMO-OFDM signals, the waveform typically exhibits a large PAPR. When the instantaneous peaks approach the saturation region of the PA, severe nonlinear distortion occurs, which degrades the error vector magnitude (EVM). To satisfy practical modulation quality constraints, the average output power must be sufficiently backed off from saturation. In this context, the required OBO is directly dictated by the signal statistics to ensure the peaks remain within the quasi-linear region. Moreover, for practical RF power amplifiers, especially Class-AB, Doherty, and related high-efficiency architectures, the drain efficiency generally decreases as the operating point moves away from saturation, which has been widely discussed in the PA literature \cite{raab2002power,camarchia2015doherty,singh2021review}.

\begin{figure}[!t]
\centering
\begin{minipage}{0.58\linewidth}
  \centering
  \includegraphics[width=\linewidth]{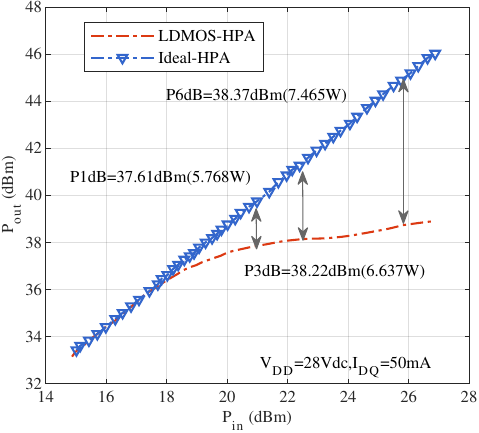}
  \vspace{2pt}
  {\footnotesize\textbf{(a)}~Ideal versus practical models.\par}
\end{minipage}

\vspace{6pt} 

\begin{minipage}{0.64\linewidth}
  \centering
  \includegraphics[width=\linewidth]{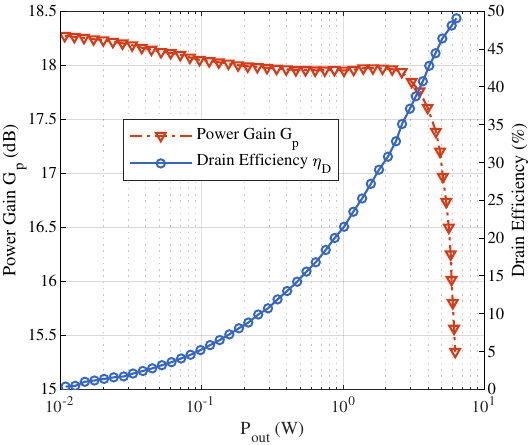}
  \vspace{2pt}
  {\footnotesize\textbf{(b)}~Gain and efficiency curves.\par}
\end{minipage}

\caption{LDMOS-based power amplifier characteristics: (a) ideal vs. practical models; (b) gain and efficiency curves.}
\label{fig:HPA_combined}
\end{figure}

To characterize the impact of waveform statistics on hardware performance, we employ the notation $\uparrow$ and $\downarrow$ to represent the monotonic increase and decrease of the corresponding physical quantities, respectively. As the OBO increases, the PA's operating point shifts further away from the high-efficiency saturation region, leading to a reduction in drain efficiency. For typical Class-AB or LDMOS amplifiers, this physical constraint establishes a fundamental qualitative relationship, as shown in Fig. \ref{fig:HPA_combined}:
\begin{equation}
\mathbb{E}[\text{PAPR}_{\text{TX}}] \uparrow \implies \text{OBO} \uparrow \implies \text{Operating Point} \downarrow \implies \eta_{\text{PA}} \downarrow.
\end{equation}

The above chain of implication illustrates that the required OBO is positively correlated with $\mathbb{E}[\text{PAPR}_{\text{TX}}]$. Consequently, reducing $\mathbb{E}[\text{PAPR}_{\text{TX}}]$ allows for a smaller OBO, thereby enabling the PA to operate at a higher efficiency point closer to saturation.

\subsection{Rectifier}

\begin{figure}[!t]
\centering

\begin{minipage}{0.9\linewidth}
  \centering
  \includegraphics[width=\linewidth]{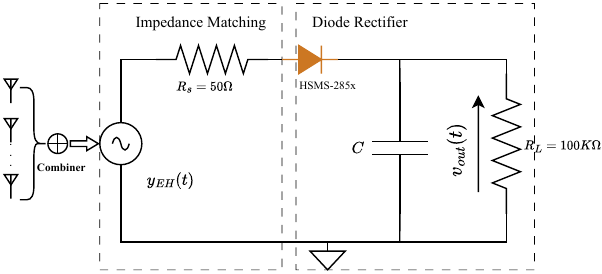}
  \vspace{2pt}
  {\footnotesize
    \parbox{0.95\linewidth}{\centering \textbf{(a)}~Schematic diagram of a diode-based rectifier circuit.}
  }
\end{minipage}

\vspace{6pt} 

\begin{minipage}{0.9\linewidth}
  \centering
  \includegraphics[width=0.64\linewidth]{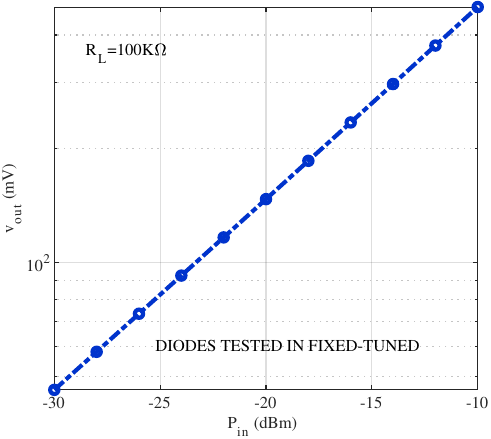}
  \vspace{2pt}
  {\footnotesize
    \parbox{0.95\linewidth}{\centering \textbf{(b)}~Rectification characteristics: input RF power vs. output DC voltage.}
  }
\end{minipage}

\caption{Diode-based rectifier characteristics: (a) circuit schematic; (b) input–output power relationship.}
\label{fig:rectifier_combined}
\end{figure}

As illustrated in Fig.~\ref{fig:rectifier_combined}(a), a typical single-diode-based rectifier circuit is considered. The corresponding RF-to-DC conversion efficiency can be expressed as~\cite{valenta2015theoretical,rotenberg2020efficient}

\begin{equation}
{\eta _{\mathrm{R}}} = \frac{{{P_{{\text{DC}},{\text{ out }}}}}}{{{P_{{\text{RF}},{\text{in}}}}}} = \frac{{{v_{out}}^2/{R_L}}}{{{P_{{\text{RF}},{\text{in}}}}}}.
\end{equation}

The rectifying diode employed in this work is the Avago HSMS-285x series~\cite{hsms285x}, which has been widely used in RF energy harvesting circuits due to its low turn-on voltage and small junction capacitance. Such Schottky devices have been adopted in a number of experimental rectifier designs, confirming their high sensitivity and suitability for low-power operation. As shown in Fig.~\ref{fig:rectifier_combined}(b), the measured relationship between the output voltage $v_{out}$ and the input RF power $P_{\text{RF,in}}$ for the HSMS-285x diode is illustrated.

The starting point for modeling is the exponential $i$--$v$ law of the diode, including reverse breakdown effects. Specifically, the instantaneous diode current can be written as~\cite{abeywickrama2021refined}
\begin{equation}
i_d(t) = I_0 \!\left(e^{\tfrac{v_d(t)}{\eta V_0}}-1\right)
- I_{BV}\,e^{-\tfrac{v_d(t)+V_B}{\eta V_0}},
\end{equation}
where $I_0$ denotes the reverse saturation current, $I_{BV}$ is the breakdown saturation current, $V_B$ is the breakdown voltage, $\eta$ is the ideality factor, and $V_0$ is the thermal voltage. A single-diode rectifier followed by a large smoothing capacitor $C$ and a resistive load $R_L$ is considered. Under the large-capacitor assumption, the average load current equals the average diode current, which yields~\cite{clerckx2016waveform}
\begin{equation}
\frac{v_{out}}{R_L} = \frac{1}{T}\int_0^T i_d(t)\,dt.
\end{equation}

Let $y(t)$ denote the normalized RF input waveform such that $\sqrt{R_s}|y(t)|$ equals the instantaneous RF voltage magnitude across the diode, with $R_s$ being the source resistance. Following the standard envelope approximation commonly used in rectifier analysis, the diode voltage can then be expressed as
\begin{equation}
v_d(t) = \sqrt{R_s}\,|y(t)| - v_{out}.
\end{equation}

Here, the absolute value models ideal full-wave rectification, which slightly overestimates current at very low input levels but is accurate in moderate-to-high input regimes.

Substituting into the current expression and averaging, two types of exponential averages appear. For compactness, we define
\begin{align}
\Phi_{+} &\triangleq \frac{1}{T}\int_0^T 
   \exp\!\left(\frac{\sqrt{R_s}|y(t)|}{\eta V_0}\right)\,dt, \\
\Phi_{-} &\triangleq \frac{1}{T}\int_0^T 
   \exp\!\left(-\frac{\sqrt{R_s}|y(t)|}{\eta V_0}\right)\,dt.
\end{align}

The averaged current balance then becomes
\begin{equation}
\frac{v_{out}}{R_L}
= I_0 \!\left( e^{-\tfrac{v_{out}}{\eta V_0}}\Phi_{+}-1 \right)
- I_{BV}\,e^{\tfrac{v_{out}-V_B}{\eta V_0}}\,\Phi_{-}.
\end{equation}

Rearranging terms gives
\begin{align}
\Phi_{+} &= e^{\tfrac{v_{out}}{\eta V_0}}
\left(1+\frac{v_{out}}{R_L I_0}\right)
+ \frac{I_{BV}}{I_0} e^{\tfrac{2v_{out}-V_B}{\eta V_0}} \Phi_{-}.
\end{align}

When breakdown must be included, a common approximation is to assume $\Phi_{+}\!\approx\!\Phi_{-}$, which is accurate when the input envelope distribution is narrow or when operation is well below breakdown. Under this assumption, the above relation reduces to
\begin{equation}
\Phi_{+} =
\frac{
\exp\!\left(\tfrac{v_{out}}{\eta V_0}\right)\left(1+\tfrac{v_{out}}{R_L I_0}\right)
}{
1 - \left(\tfrac{I_{BV}}{I_0}\right)\exp\!\left(\tfrac{2v_{out}-V_B}{\eta V_0}\right)
}.
\end{equation}

This approximation is widely used for analytical tractability, while more refined rectifier models that capture device-level nonlinearities have been proposed in recent work.

Finally, by substituting the definition of $\Phi_{+}$, the rectifier equation can be expressed as
\begin{equation}
\label{eq:rectifier_equation_final}
\frac{
    \exp \!\left(\tfrac{v_{out}}{\eta V_0}\right)
    \left(1 + \tfrac{v_{out}}{R_L I_0}\right)
}{
    1 - \left(\tfrac{I_{BV}}{I_0}\right)
    \exp \!\left(\tfrac{2v_{out}-V_B}{\eta V_0}\right)
}
= \frac{1}{T}\int_0^T 
    \exp \!\left(\tfrac{\sqrt{R_s}\,|y(t)|}{\eta V_0}\right)\,dt .
\end{equation}

It can be observed that the left-hand side of~\eqref{eq:rectifier_equation_final} is a monotonically non-decreasing function of the output voltage $v_{out}$, while the right-hand side depends on the statistics of the input waveform $y(t)$~\cite{abeywickrama2021refined}. As a result, the rectifier output voltage, and hence the RF-to-DC conversion efficiency, is positively correlated with the exponential moment of the received signal envelope. This relation can be written as
\begin{equation}
{\eta _{\mathrm{R}}} \uparrow \iff v_{out} \uparrow \iff \frac{1}{T}\int_0^T e^{\alpha |y(t)|} dt \uparrow,
\label{eq6}
\end{equation}
where $\alpha  = \frac{\sqrt{R_s}}{\eta V_0}$. Approximating the integral by discretization yields
\begin{equation}
\begin{aligned}
\eta_{\mathrm{R}} \uparrow & \iff \frac{1}{N} \sum_{n=1}^{N} e^{\alpha|y[n]|} \uparrow
\approx \exp \!\big(\alpha \max \{|y[n]|\}\big) \uparrow.
\end{aligned}
\label{eq7}
\end{equation}
where the approximation follows from the dominance of waveform peaks in the nonlinear exponential model of rectification. 

Therefore, a key conclusion is obtained: the RF-to-DC conversion efficiency of the rectifier increases with the peak amplitude of the received signal, and is hence positively correlated with the PAPR of the input waveform. This fundamental property has been confirmed by both circuit-level simulations and experimental validations in prior works~\cite{clerckx2016waveform,valenta2015theoretical,collado2014optimal} as
\begin{equation}
\ln(\eta_{\mathrm{R}}) \uparrow \iff \max \{|y[n]|\} \uparrow \iff \mathbb{E}\!\left[\text{PAPR}_{\text{RX}}\right] \uparrow.
\end{equation}

\subsection{Modeling and Analysis of Energy Efficiency}

Based on the preceding analysis, the end-to-end efficiency considered in this
work is jointly determined by the PA efficiency and the rectifier efficiency:
\begin{equation}
\eta = \eta_{\mathrm{PA}} \eta_{\mathrm{R}} \uparrow \iff
\frac{\mathbb{E}\!\left[\mathrm{PAPR}_{\mathrm{RX}}\right]}
{\mathbb{E}\!\left[\mathrm{PAPR}_{\mathrm{TX}}\right]} \uparrow,
\end{equation}
where the above positive correlation holds under a fixed peak-envelope
constraint (or a fixed nonlinear-distortion level) and moderate operating
conditions for both the PA and the rectifier.

Under a fixed peak power constraint $P_{\text{sat}}$, a lower transmit-side
PAPR allows the average transmit power $P_{\text{out}}$ to increase without
violating the peak limit or increasing nonlinear distortion. This reduces the
required OBO, moves the PA closer to its high-efficiency saturation region
(Fig.~\ref{fig:HPA_combined}), and simultaneously improves the received SNR.
Conversely, a higher receive-side PAPR increases the instantaneous peak
amplitude of the RF waveform, strengthens the nonlinear rectification
process, and thus improves the RF-to-DC conversion efficiency
$\eta_{\mathrm{R}}$.

Unlike conventional waveform designs that mainly reduce transmit-side PAPR,
this work explicitly exploits the dual role of PAPR in SWIPT: a lower
$\mathbb{E}[\mathrm{PAPR}_{\mathrm{TX}}]$ benefits PA efficiency, while a
higher $\mathbb{E}[\mathrm{PAPR}_{\mathrm{RX}}]$ benefits rectification. To
capture this combined effect, we introduce the following dimensionless
metric:
\begin{equation}
\Xi =
\frac{\mathbb{E}\!\left[\mathrm{PAPR}_{\mathrm{RX}}\right]}
{\mathbb{E}\!\left[\mathrm{PAPR}_{\mathrm{TX}}\right]},
\label{papr_metric}
\end{equation}
where the expectation is taken over OFDM symbols and channel realizations.

The rationale is summarized by the qualitative trends
$\mathbb{E}[\mathrm{PAPR}_{\mathrm{TX}}] \uparrow \Rightarrow \mathrm{OBO}
\uparrow \Rightarrow \eta_{\mathrm{PA}} \downarrow$ and
$\mathbb{E}[\mathrm{PAPR}_{\mathrm{RX}}] \uparrow \Rightarrow
\eta_{\mathrm{R}} \uparrow$ within the practical operating regime. Here, the
PA is assumed to operate in the moderate back-off region, while the
rectifier is governed by the nonlinear diode response in the regime where its
dominant nonlinear behavior remains active. The impact of channel multipath
and spatial combining is reflected through the received-waveform statistics
and hence in $\mathbb{E}[\mathrm{PAPR}_{\mathrm{RX}}]$. Therefore, $\Xi$
should be interpreted as a design-oriented, equivalent-channel
performance indicator, for which an approximately monotonic correlation with
the end-to-end efficiency is observed in the practical operating regime.
Outside this regime, such as under deep PA saturation, rectifier breakdown,
or severe mismatch of the equivalent-channel abstraction, $\Xi$ should be
viewed as a useful design indicator rather than a strict physical law.

In the proposed framework, $\Xi$ is not used as a standalone optimization
objective; instead, it serves as the system-level design rationale
for the subsequent hierarchical optimization. Specifically, the allocation
variables of subcarriers determine the tradeoff among
receive-side peak enhancement, transmit-side peak-reduction capability, and
information transmission, while the waveform variable $X_{\mathrm{TR}}$ is
optimized subsequently to realize transmit-side PAPR suppression under the
obtained allocation. Hence, the overall method is a coherent two-layer
design under the same system model, with $\Xi$ providing the global guiding
principle.

\section{Proposed System Architecture}

In this work, a power-splitting receiver architecture is adopted~\cite{zhang2013mimo,zhou2013wireless,clerckx2018fundamentals}, where the received signal is divided into two branches for simultaneous information decoding (ID) and energy harvesting (EH). To enhance this baseline model, a frequency-domain resource allocation scheme is incorporated. Specifically, the available spectrum is partitioned into three functional segments: a PAPR-controlled block employed to improve PA efficiency, a low-rate block shaped to enhance waveform peaks and facilitate rectification, and a high-rate block dedicated to sustaining conventional data transmission. By adaptively adjusting the allocation across these segments, a flexible trade-off between energy transfer efficiency and spectral efficiency can be achieved.

For an MIMO-OFDM system with $N_t$ transmit antennas, $N_r$ receive antennas, $N_s$ spatial streams, and $K$ subcarriers, the transmitted signal is represented by the matrix ${\mathbf{X}} \in {\mathbb{C}^{{N_t} \times K}}$, where each row corresponds to the frequency-domain symbols transmitted by one antenna over $K$ subcarriers, i.e.,

\begin{equation}
\begin{array}{*{20}{l}}
  {{\mathbf{X}} = [\underbrace {{{\mathbf{X}}^{{\text{TR}}}}}_{{\text{PAPR Reduction}}}{\text{ }}\underbrace {{\mathbf{W}} \odot {{\mathbf{X}}^{{\text{Infor}}}}}_{{\text{Information Transfer }}}]},
\end{array}
\end{equation}
where ${{\mathbf{X}}^{{\text{TR}}}} \in {\mathbb{C}^{{N_t} \times {K_{{\text{TR}}}}}}$ represents the TR symbols employed for PAPR reduction, ${\mathbf{W}} \in {\mathbb{C}^{{N_t} \times {N_s} \times K}}$ denotes the precoding matrix, and $\odot$ indicates subcarrier-wise matrix multiplication. The information-bearing component ${{\mathbf{X}}^{{\text{Infor}}}} \in {\mathbb{C}^{{N_s} \times ({K_{{\text{QAM}}}} + {K_{{\text{IM}}}})}}$ is further divided into two parts to enable simultaneous index-modulated diversity transmission and spatially multiplexed QAM transmission, as

\begin{align}
\mathbf{X}^{\text {Infor }} & = [\underbrace{\mathbf{X}^{\mathrm{IM}}}_{\text {Spatial Diversity}}
\underbrace{\mathbf{X}^{\mathrm{QAM}}}_{\text {Spatial Multiplexing }}],
\end{align}
where ${{\mathbf{X}}^{{\text{IM}}}} \in {\mathbb{C}^{{N_s} \times {K_{{\text{IM}}}}}}$ applies index modulation (IM) to enhance rectification efficiency at the receiver, while ${{\mathbf{X}}^{{\text{QAM}}}} \in {\mathbb{C}^{{N_s} \times {K_{{\text{QAM}}}}}}$ carries quadrature amplitude modulation (QAM) symbols to maintain high spectral efficiency. Let $\mathcal{K}$ denote the index set of all subcarriers, and let
$\mathcal{K}_{\mathrm{TR}}$, $\mathcal{K}_{\mathrm{IM}}$, and
$\mathcal{K}_{\mathrm{QAM}}$ denote the subsets assigned to tone reservation
(TR), index modulation (IM), and conventional QAM transmission, respectively.
These three subsets are mutually disjoint and satisfy
\begin{equation}
\begin{aligned}
\mathcal{K}
&=
\mathcal{K}_{\mathrm{TR}}
\cup
\mathcal{K}_{\mathrm{IM}}
\cup
\mathcal{K}_{\mathrm{QAM}},\\
\mathcal{K}_{\mathrm{TR}} \cap \mathcal{K}_{\mathrm{IM}}
&=
\mathcal{K}_{\mathrm{TR}} \cap \mathcal{K}_{\mathrm{QAM}}
=
\mathcal{K}_{\mathrm{IM}} \cap \mathcal{K}_{\mathrm{QAM}}
=
\varnothing.
\end{aligned}
\end{equation}

Their cardinalities are denoted by
$K_{\mathrm{TR}}=|\mathcal{K}_{\mathrm{TR}}|$,
$K_{\mathrm{IM}}=|\mathcal{K}_{\mathrm{IM}}|$, and
$K_{\mathrm{QAM}}=|\mathcal{K}_{\mathrm{QAM}}|$, respectively, and therefore
\begin{equation}
K = K_{\mathrm{TR}} + K_{\mathrm{IM}} + K_{\mathrm{QAM}}.
\end{equation}

At the receiver, the frequency-domain signals from all antennas are stacked into a matrix ${{\mathbf{Y}}} \in {\mathbb{C}^{{N_r} \times K}}$. For the $k$-th subcarrier, the received signal vector is given by
\begin{equation}
\begin{array}{*{20}{l}}
  {{{\mathbf{Y}}_k}}&{ = {{\mathbf{H}}_k}{{\widetilde {\mathbf{X}}}_k} + {{\mathbf{N}}_k}}, 
\end{array}
\end{equation}
where ${{\mathbf{H}}_k} \in {\mathbb{C}^{{N_r} \times {N_t}}}$ denotes the MIMO channel matrix, ${{\mathbf{N}}_k}$ is the additive white Gaussian noise (AWGN) vector with ${\mathbb{E}\left[ {{{\mathbf{N}}_k}{\mathbf{N}}_k^H} \right] = \sigma _n^2{{\mathbf{I}}_{{N_r}}}}$, and
\[{\widetilde {\mathbf{X}}_k} = {{\mathbf{W}}_k}{{\mathbf{X}}_k},{{\widetilde {\mathbf{X}}_k}} \in {\mathbb{C}^{{N_t} \times K}},{{\mathbf{W}}_k} \in {\mathbb{C}^{{N_t} \times {N_s}}}.\]

The received time-domain signal at the $i$-th antenna is expressed as
\begin{equation}
{{y_{i,k}} = \frac{1}{{\sqrt K }}\sum\limits_{k = 0}^{{K} - 1} {{Y_{i,k}}} {e^{j2\pi kn/K}},\quad n = 0, \ldots ,K - 1},
\end{equation}
where ${{Y_{i,k}}}$ denotes the frequency-domain symbol for the $k$-th subcarrier at the $i$-th receive antenna. The corresponding time-domain signal vector is then written as
\begin{equation}
\begin{array}{*{20}{l}}
  {{{\mathbf{y}}_i} = {{\left[ {{y_i}[1],{y_i}[2], \ldots ,{y_i}[K]} \right]}^T} \in {\mathbb{C}^{1 \times K}}}.
\end{array}
\end{equation}

Given a power splitting ratio $\rho \in (0,1)$, the received signals are
split into two components for simultaneous information decoding and
energy harvesting. Let
\[
\mathbf{Y}
=
\begin{bmatrix}
\mathbf{y}_1 \\
\mathbf{y}_2 \\
\vdots \\
\mathbf{y}_{N_r}
\end{bmatrix}
\in \mathbb{C}^{N_r \times K}
\]
denote the received signal matrix formed by stacking the $K$-sample received
vectors from all $N_r$ antennas. Then the ID and EH signals are expressed as
\begin{equation}
\left\{
\begin{aligned}
\mathbf{Y}_{\mathrm{ID}} &=
\sqrt{\rho}\,\mathbf{Y}
\in \mathbb{C}^{N_r \times K},\\
\mathbf{y}_{\mathrm{EH}} &=
\sum_{i=1}^{N_r}\sqrt{1-\rho}\,\mathbf{y}_i
\in \mathbb{C}^{1 \times K},
\end{aligned}
\right.
\end{equation}
where $\mathbf{y}_{\mathrm{EH}}$ is obtained by combining the received signals
from multiple antennas.

\begin{figure}[!t]
\centering

\includegraphics[width=8cm]{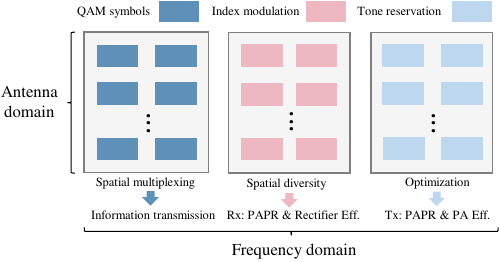}
\caption{Subcarrier allocation strategy in the proposed SWIPT waveform design.}
\label{fig6}
\end{figure}
\subsection{High PAPR for Rectifier via Index Modulation}

To enhance the receive-side PAPR in a controllable manner, we introduce a
specialized IM waveform component, denoted by $X^{\mathrm{IM}}$, whose role is
not merely to convey index information, but more importantly to shape
the received time-domain waveform so as to create a dominant peak at the
energy-harvesting branch. This design is motivated by the fact that a peaky
received RF waveform is beneficial for exciting the rectifier nonlinearity and
thus improving the RF-to-DC conversion efficiency.

Specifically, the adopted IM mapping is defined as
\begin{equation}
\begin{array}{*{20}{c}}
  \text{Bits} & \text{Subblocks} \\ 
  0 & {[s_\chi \ \ 0]^T} \\ 
  1 & {[0 \ \ s_\chi]^T},
\end{array}
\end{equation}
where only two symbols are employed, namely $0$ and a single non-zero symbol
$s_\chi = A_0 e^{j\theta_0}$, where $\theta_0$ is a fixed design phase chosen
so that the non-zero IM components are phase-aligned and coherently combined
at the target time-domain sample. For signal-power normalization, the
amplitude is set to $A_0 = \sqrt{2/N_r}$. Unlike a conventional comb-type allocation, where
different functional subcarriers are randomly interleaved, the proposed
structure is chosen to provide a clearer functional separation among the three
design objectives: the IM component is used to create a strong received peak,
the TR component is used to suppress the transmit-side PAPR, and the remaining
QAM component is used to preserve information transmission. This separation
makes the receive-side peak formation more controllable and facilitates the
joint balancing of rectifier-oriented peak enhancement and transmitter-side
peak reduction.

By the Fourier transform property, the in-phase IM components are coherently
superimposed at the first time-domain sample. As a result, the first sample of
the harvested-signal waveform becomes the dominant peak candidate, i.e.,
\begin{equation}
|{\mathbf{y}}_{\text{EH}}[0]|
=
\left|
\sqrt{1-\rho}\sum_{i=1}^{N_r}\sum_{k=0}^{K-1} Y_{i,k}
\right|.
\end{equation}

In the proposed construction, this sample is intentionally designed to carry
the strongest coherent contribution and is therefore used as the dominant peak
for the following receive-side PAPR analysis.

For standard QAM constellations, the quadrantal symmetry in the complex plane
implies that the transmitted symbols are zero-mean. Hence, the QAM-modulated
components satisfy
\begin{equation}
\mathbb{E}\!\left[
\sum_{k \in \mathcal{K}_{\text{QAM}}} Y^{\text{QAM}}_{i,k}
\right] = 0.
\end{equation}

In contrast, the IM-modulated symbols exhibit a deterministic phase-aligned
bias. The expected coherent superposition of the IM components at the first
time sample is
\begin{equation}
\mathbb{E}\!\left[
\sum_{k \in \mathcal{K}_{\text{IM}}} Y^{\text{IM}}_{i,k}
\right]
=
\frac{K_{\text{IM}}}{\sqrt{2N_r}} e^{j\theta_0}.
\label{eq20}
\end{equation}

This expression shows that the IM component contributes a non-zero coherent term
to the received waveform, which is precisely why $X^{\mathrm{IM}}$ is
introduced in the proposed design.

To counterbalance the transmit-side peak growth caused by this coherent
superposition, the TR subcarriers are designed to generate an opposite peak at
the same time-domain position, thereby suppressing the transmit-side PAPR.
This cancellation effect persists after wireless propagation, and its average
contribution is modeled as
\begin{equation}
\mathbb{E}\!\left[
\sum_{k \in \mathcal{K}_{\text{TR}}} Y_{i,k}^{\text{TR}}
\right]
=
-\frac{\beta K_{\text{TR}}}{\sqrt{N_r}} e^{j\theta_0},
\label{eq21}
\end{equation}
where $\beta$ denotes a small scaling factor that characterizes the average
peak-cancellation effect of the TR component.

Combining (\ref{eq20}) and (\ref{eq21}), and using the fact that the QAM part
has zero mean, the coherent component at the dominant first time sample can be
approximated as
\begin{equation}
\mathbb{E}\!\left[{\mathbf{y}}_{\text{EH}}[0]\right]
\approx
\left(\frac{K_{\text{IM}}}{\sqrt{2}}-\beta K_{\text{TR}}\right)\sqrt{N_r}.
\end{equation}

Accordingly, under the proposed waveform construction, where the first
time-domain sample is designed to be the dominant peak, we adopt the following
dominant-peak approximation:
\begin{equation}
\mathbb{E}\!\left[\max \big\{|{\mathbf{y}}_{\text{EH}}|\big\}\right]
\approx
\left(\frac{K_{\text{IM}}}{\sqrt{2}}-\beta K_{\text{TR}}\right)\sqrt{N_r}.
\label{eq:dominant_peak_approx}
\end{equation}

We emphasize that this is a design-oriented approximation based on the
engineered coherent superposition at the first time sample, rather than a
general identity replacing the expectation of the maximum by the maximum of the
expectation.

Consequently, the receive-side PAPR after power combining is approximated as
\begin{equation}
\begin{aligned}
  \mathbb{E}\!\left[\text{PAPR}_{{\mathbf{y}}_{\text{EH}}}\right]
  &=
  10\log_{10}\!\left(
  \frac{\mathbb{E}\!\left[\max\{|{\mathbf{y}}_{\text{EH}}|^2\}\right]}
  {\mathbb{E}[|{\mathbf{y}}_{\text{EH}}|^2]}
  \right) \\
  &\approx
  20\log_{10}\!\left(
  \frac{K_{\text{IM}}}{\sqrt{2}}-\beta K_{\text{TR}}
  \right)
  + 10\log_{10}(N_r),
\label{eq24}
\end{aligned}
\end{equation}
where the approximation follows from (\ref{eq:dominant_peak_approx}) and from
the normalization of the average received signal power. Therefore, increasing
$K_{\text{IM}}$ strengthens the coherent peak formation at the receiver,
whereas increasing $K_{\text{TR}}$ counteracts this peak to control the
transmit-side PAPR, revealing the fundamental tradeoff between receiver-side
rectifier enhancement and transmitter-side PA efficiency.

As demonstrated in the previous section, the proposed scheme substantially
increases the value of $\Xi$ in (\ref{papr_metric}) by enhancing the
receive-side PAPR, thereby yielding a significant improvement in rectification
efficiency.

\begin{figure*}[!t]
\centering
\includegraphics[width=18cm]{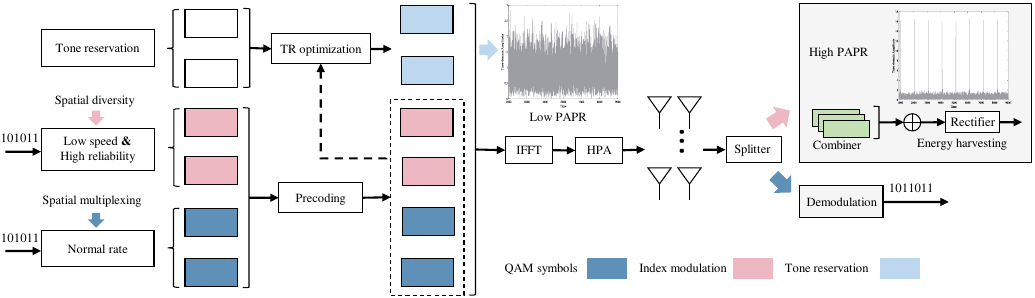}
\caption{System block diagram of the proposed SWIPT scheme.}
\label{fig_1}
\end{figure*}

\subsection{PAPR Reduction for Amplifier via Tone Reservation}
After the mapping of ${{\mathbf{X}}^{\text{Infor}}}$ is completed, PAPR reduction is performed by optimizing the symbols placed on the reserved subcarriers. In this work, a TR algorithm~\cite{li2018tone} with accelerated convergence based on gradient descent is employed. For the $i$-th transmit antenna, any row of ${\mathbf{X}} \in \mathbb{C}^{N_t \times K}$ corresponds to all frequency-domain symbols, expressed as  

\[
{{\mathbf{x}}_i} = [{{\mathbf{x}}_i}[1], {{\mathbf{x}}_i}[2], \ldots, {{\mathbf{x}}_i}[K]], 
\quad {{\mathbf{x}}_i} \in \mathbb{C}^{1 \times K}.
\]

The peak power of the corresponding time-domain signal is then written as  

\[
f({\mathbf{c}}_i) = \max \left\{ {\mathbf{F}}^{\text{H}}({\mathbf{x}}_i + {\mathbf{c}}_i) \right\} 
= \max \{ {\mathbf{z}} \},
\]
where ${\mathbf{F}}$ denotes the Fourier transform matrix, ${\mathbf{c}}_i \in \mathbb{C}^{1 \times K}$ represents the TR signal for PAPR reduction in the frequency domain of the $i$-th antenna, and  

\[
{\mathbf{z}} = {\mathbf{F}}^{\text{H}}({\mathbf{x}}_i + {\mathbf{c}}_i).
\]

Thus, for each antenna, the following optimization problem is formulated:
\begin{align*} 
  \mathop {\min } \limits_{{{\mathbf{c}}_i}} \ & f({{\mathbf{c}}_i}) \hfill \\
  \text{s.t.} \quad {{\mathbf{c}}_i}[k] &= 0, \quad k \in \mathcal{K}_{\text{IM}} \cup \mathcal{K}_{\text{QAM}} \hfill \\
  \left\| {{{\mathbf{c}}_i}} \right\|_{\text{F}}^2 &< \frac{1}{2}\frac{K_{\text{TR}}}{K - K_{\text{TR}}}\left\| {{{\mathbf{x}}_i}} \right\|_{\text{F}}^2. \hfill
\end{align*}

Since the objective function is convex owing to the $\max\{\cdot\}$ operator, the optimization problem can be efficiently solved using convex optimization tools such as the CVX toolbox. The detailed steps are summarized in \textbf{Algorithm 1}.

\begin{algorithm}[t]
\caption{CVX-Based Tone Reservation Algorithm for Multi-Antenna PAPR Reduction}
\label{alg:TR_CVX_multi}
\begin{algorithmic}[1]
\State \textbf{Input:} number of transmit antennas $N_t$; data symbol vectors $\{\mathbf{x}_i\}_{i=1}^{N_t}$; reserved-tone index set $\mathcal{K}_{\text{TR}}$; IFFT size $K$
\State \textbf{Output:} time-domain OFDM signals $\{\mathbf{s}_i\}_{i=1}^{N_t}$ with reduced PAPR
\For{$i=1$ to $N_t$}
    \State Initialize reserved-tone vector $\mathbf{c}_i$ supported on $\mathcal{K}_{\text{TR}}$
    \State Construct frequency-domain signal $\widetilde{\mathbf{x}}_i = \mathbf{x}_i + \mathbf{c}_i$
    \State Let $\mathbf{F}^H$ denote the $K$-point IFFT matrix
    \State Solve the convex optimization problem via \textbf{CVX}:
    \begin{align*}
    \min_{\mathbf{c}_i} \quad & \big\| \mathbf{F}^H (\mathbf{x}_i + \mathbf{c}_i) \big\|_{\infty} \\
    \text{s.t.} \quad & \mathbf{c}_i[k] = 0, \quad k \notin \mathcal{K}_{\text{TR}}, \\
    & \| \mathbf{c}_i \|_{\text{F}}^2 < \tfrac{1}{2}\tfrac{K_{\text{TR}}}{K - K_{\text{TR}}}\, \| \mathbf{x}_i \|_{\text{F}}^2
    \end{align*}
    \State Compute time-domain signal: $\mathbf{s}_i \gets \mathbf{F}^{H} \big(\mathbf{x}_i + \mathbf{c}_i\big)$
\EndFor
\State \textbf{Return} $\{\mathbf{s}_i\}_{i=1}^{N_t}$ and $\{\widetilde{\mathbf{x}}_i\}_{i=1}^{N_t}$
\end{algorithmic}
\end{algorithm}

However, due to the high computational complexity of the convex optimization approach, a gradient descent-based TR algorithm is developed in this paper. In this method, the $\max \{ \cdot \}$ function is approximated by a smooth logarithmic-exponential form, which facilitates efficient and fast optimization of the reserved subcarrier symbols. The detailed derivation is presented as follows.

First, the objective function is approximated as
\[
\tilde f({\mathbf{c}}_i) = \frac{1}{\varepsilon }\ln \!\left(\sum\limits_{k = 1}^K e^{\varepsilon z_k}\right) 
\approx \max \{ {\mathbf{z}}\} = f({\mathbf{c}}_i),
\]
where $\varepsilon$ is a smoothing parameter.  

The partial gradient ${\mathbf{g}} \in \mathbb{C}^{1 \times K}$ with respect to the TR subcarriers can be expressed as
\begin{equation}
{\mathbf{g}} = \nabla \tilde f({\mathbf{c}}_i) = 
\begin{cases}
  {\mathbf{\tilde g}}[k], & k \in \mathcal{K}_{\text{TR}}, \\[6pt]
  0, & k \in \mathcal{K}_{\text{IM}} \cup \mathcal{K}_{\text{QAM}},
\end{cases}
\label{gradient1}
\end{equation}
where
\begin{equation}
{\mathbf{\tilde g}} 
= {\mathbf{F}} \nabla f({\mathbf{z}}) 
= \frac{{\mathbf{F}}}{\sum\limits_{i = 1}^K e^{\varepsilon z_i}}
\left[e^{\varepsilon z_1}, e^{\varepsilon z_2}, \ldots, e^{\varepsilon z_K}\right].
\label{gradient2}
\end{equation}

Thus, the gradient of the smoothed objective function is obtained. Based on this result, the final TR symbols are iteratively updated using the gradient descent procedure, as summarized in \textbf{Algorithm 2}. In this way, the transmit-side PAPR is effectively reduced, corresponding to the denominator in the definition of $\Xi$ in (\ref{papr_metric}), which further enhances overall energy transfer efficiency.

\begin{algorithm}[t]
\caption{Tone Reservation Algorithm for PAPR Reduction (Multi-Antenna)}
\label{alg:TR_multi}
\begin{algorithmic}[1]
\State \textbf{Input:} number of transmit antennas $N_t$; data symbol vectors $\{\mathbf{x}_i\}_{i=1}^{N_t}$; index set of reserved tones $\mathcal{K}_{\text{TR}}$; number of IFFT points $K$; maximum iterations $I_{\max}$; gradient step size $\alpha_{\mathrm{TR}}$; minimum gradient norm threshold $\varepsilon$
\State \textbf{Output:} time-domain OFDM signals $\{\mathbf{s}_i\}_{i=1}^{N_t}$ with reduced PAPR
\For{$i=1$ to $N_t$}
    \State Initialize reserved tones $\mathbf{c}_i[k]$ for $k \in \mathcal{K}_{\text{TR}}$ with random complex values and normalize them to satisfy the power constraint
    \State Construct the full frequency-domain signal $\widetilde{\mathbf{x}}_i = \mathbf{x}_i + \mathbf{c}_i$, with $\mathbf{c}_i[k] = 0$ for $k \in \mathcal{K}_{\text{IM}} \cup \mathcal{K}_{\text{QAM}}$
    \State Set the iteration counter $n \gets 0$
    \While{$n < I_{\max}$ \textbf{and} $\lVert \mathbf{g} \rVert_{\mathrm{F}} > \varepsilon$}
        \State Compute the time-domain signal: $\mathbf{z} = \mathbf{F}^{\mathrm{H}} \widetilde{\mathbf{x}}_i$
        \State Compute the gradient using (\ref{gradient1}) and (\ref{gradient2})
        \State Update the reserved-tone vector: $\mathbf{c}_i \gets \mathbf{c}_i - \alpha_{\mathrm{TR}} \mathbf{g}$
        \State Update the full frequency-domain signal: $\widetilde{\mathbf{x}}_i = \mathbf{x}_i + \mathbf{c}_i$
        \State Update $n \gets n+1$
    \EndWhile
    \State $\mathbf{s}_i \gets \mathbf{F}^{\mathrm{H}} \widetilde{\mathbf{x}}_i$
\EndFor
\State \textbf{Return} $\{\mathbf{s}_i\}_{i=1}^{N_t}$, $\{\widetilde{\mathbf{x}}_i\}_{i=1}^{N_t}$
\end{algorithmic}
\end{algorithm}

The reserved tones in Step 4 are initialized with small-magnitude 
random complex values to avoid bias toward any specific peak pattern. 
Since the tone reservation problem is nonconvex after smoothing, 
different initializations may, in principle, lead to different local 
solutions. However, extensive numerical experiments indicate that the proposed 
gradient-based procedure exhibits stable convergence behavior. 
The final PAPR reduction varies marginally across different random 
initializations, and the performance gap remains within 0.1–0.2 dB 
in the considered scenarios. This suggests that the solution landscape 
is relatively well-conditioned under the adopted smoothing strategy.

\subsection{R–E Region Characterization and Tradeoff Analysis}

In SWIPT systems, the rate--energy (R--E) region characterizes the
fundamental tradeoff between the achievable information transmission rate and
the harvested energy. It is defined as the set of all rate--energy pairs that
can be simultaneously supported under a given transmission strategy and
resource constraint, expressed as
\begin{equation}
\mathcal{R}_{\mathrm{RE}}
=
\bigcup_{\rho \in [0,1]}
\left\{
(R,E)\,\middle|\,
R \leq R(\rho),\;
E \leq E(\rho)
\right\},
\end{equation}
where $R(\cdot)$ denotes the achievable-rate related objective and
$E(\cdot)$ denotes the harvested-power function.

In this work, the transmit-side power budget is assumed to be fixed, and the
effectively radiated RF power is therefore determined by the PA efficiency.
For information transmission, both the PA efficiency and the power-splitting
ratio $\rho$ affect the received signal quality after power splitting. To
obtain a tractable upper-bound expression for the achievable rate, the
MIMO-OFDM link is represented by $N_s \times K$ parallel subchannels and
characterized by an equivalent scalar channel gain. Accordingly, the
achievable-rate bound is expressed as
\begin{equation}
R(\rho,\eta_{\text{PA}})
=
N_s\!\left(K-\mu K_{\text{IM}}-K_{\text{TR}}\right)
\log_2\!\left(
1+\frac{\rho\,\eta_{\text{PA}} P G_{\text{eff}}}{\sigma_n^2}
\right),
\end{equation}
where $P$ denotes the transmit-side power budget, $\sigma_n^2$ is the noise
power at the receiver, and $G_{\text{eff}}$ denotes the effective channel
power gain that summarizes the average equivalent gain of the parallel
MIMO-OFDM subchannels after precoding/combining. The factor $\mu$ accounts
for the information contribution of the spatially diverse index-modulated
subcarriers, which also convey a limited number of bits, and is given by
\begin{equation}
\mu = 1 - \frac{1}{2N_s \cdot 2^M},
\end{equation}
where $M$ denotes the QAM modulation order. In practice, $M$ is selected
according to the supported spectral efficiency, and is approximated as
\begin{equation}
M
=
\left\lfloor
\log_2\!\left(
1+\frac{\rho\,\eta_{\text{PA}} P G_{\text{eff}}}{\sigma_n^2}
\right)
\right\rfloor.
\end{equation}

For energy harvesting, the harvested power is jointly determined by the PA
efficiency, the rectifier efficiency, and the power-splitting ratio. Under
the same equivalent-channel abstraction, the harvested-power bound is written
as
\begin{equation}
E(\rho,\eta_{\text{PA}},\eta_{\mathrm{R}})
=
\eta_{\text{PA}}\eta_{\mathrm{R}}(1-\rho)P G_{\text{eff}}.
\end{equation}

To obtain a tractable subcarrier-allocation rule while preserving the local
dependence of the PA and rectifier efficiencies on the allocation variables,
we adopt a continuous relaxation of the discrete pair
$\mathbf{k}=[K_{\mathrm{TR}},K_{\mathrm{IM}}]^T$ and construct a locally
calibrated first-order surrogate around a nominal operating point
$\mathbf{k}^{(0)}=[K_{\mathrm{TR}}^{(0)},K_{\mathrm{IM}}^{(0)}]^T$.
As discussed in Section II, both $\eta_{\mathrm{PA}}$ and $\eta_{\mathrm{R}}$
depend on the transmit- and receive-side PAPR, which in turn vary with the
subcarrier allocation and are generally nonlinear. Therefore, within a local
neighborhood of $\mathbf{k}^{(0)}$, we approximate each efficiency function
by its first-order expansion as
\begin{equation}
\eta(\mathbf{k})
\approx
\eta(\mathbf{k}^{(0)})
+
\nabla_{\mathbf{k}}\eta(\mathbf{k}^{(0)})^{T}
(\mathbf{k}-\mathbf{k}^{(0)}).
\end{equation}

Applying this approximation to the PA and rectifier efficiencies yields
\begin{equation}
\begin{aligned}
\eta_{\mathrm{PA}}
&\approx
\eta_{\mathrm{PA}}^{(0)}
+a_{\mathrm{PA}}(K_{\mathrm{TR}}-K_{\mathrm{TR}}^{(0)})
+b_{\mathrm{PA}}(K_{\mathrm{IM}}-K_{\mathrm{IM}}^{(0)}), \\
\eta_{\mathrm{R}}
&\approx
\eta_{\mathrm{R}}^{(0)}
+a_{\mathrm{R}}(K_{\mathrm{TR}}-K_{\mathrm{TR}}^{(0)})
+b_{\mathrm{R}}(K_{\mathrm{IM}}-K_{\mathrm{IM}}^{(0)}),
\end{aligned}
\end{equation}
where
\begin{equation}
\begin{aligned}
a_{\mathrm{PA}} &= \left.\frac{\partial \eta_{\mathrm{PA}}}{\partial K_{\mathrm{TR}}}\right|_{\mathbf{k}^{(0)}},
&
b_{\mathrm{PA}} &= \left.\frac{\partial \eta_{\mathrm{PA}}}{\partial K_{\mathrm{IM}}}\right|_{\mathbf{k}^{(0)}},\\
a_{\mathrm{R}} &= \left.\frac{\partial \eta_{\mathrm{R}}}{\partial K_{\mathrm{TR}}}\right|_{\mathbf{k}^{(0)}},
&
b_{\mathrm{R}} &= \left.\frac{\partial \eta_{\mathrm{R}}}{\partial K_{\mathrm{IM}}}\right|_{\mathbf{k}^{(0)}}.
\end{aligned}
\end{equation}

For compact notation, the above local models are equivalently rewritten as
\begin{equation}
\begin{aligned}
\eta_{\mathrm{PA}} &\approx C_1 (K_{\mathrm{TR}}-\beta_1 K_{\mathrm{IM}})+C_2, \\
\eta_{\mathrm{R}}  &\approx C_3 (K_{\mathrm{IM}}-\beta_2 K_{\mathrm{TR}})+C_4,
\end{aligned}
\end{equation}
with
\begin{equation}
\begin{aligned}
C_1 &= a_{\mathrm{PA}}, \qquad
\beta_1 = -\frac{b_{\mathrm{PA}}}{a_{\mathrm{PA}}},\\
C_2 &= \eta_{\mathrm{PA}}^{(0)}
-a_{\mathrm{PA}}K_{\mathrm{TR}}^{(0)}
-b_{\mathrm{PA}}K_{\mathrm{IM}}^{(0)},\\
C_3 &= b_{\mathrm{R}}, \qquad
\beta_2 = -\frac{a_{\mathrm{R}}}{b_{\mathrm{R}}},\\
C_4 &= \eta_{\mathrm{R}}^{(0)}
-a_{\mathrm{R}}K_{\mathrm{TR}}^{(0)}
-b_{\mathrm{R}}K_{\mathrm{IM}}^{(0)}.
\end{aligned}
\end{equation}

The coefficients are obtained numerically from local finite-difference
evaluation of the PA and rectifier characteristics around
$\mathbf{k}^{(0)}$, using the operating region in which the corresponding
efficiency curves are approximately linear. Hence, (34) and (35) should be
interpreted as local affine surrogate models rather than globally valid
closed-form laws. In the proposed optimization, this operating point is
updated iteratively so that the surrogate remains consistent with the current
allocation.

\begin{figure*}[hb] 
\hrulefill
\centering
\begin{equation}
{R(\rho ,{K_{{\text{TR}}}},{K_{{\text{IM}}}}) = {N_s}(K - \mu {K_{{\text{IM}}}} - {K_{{\text{TR}}}}){{\log }_2}\left( {1 + \frac{{\rho ({C_1}({K_{{\text{TR}}}} - {\beta _1}{K_{{\text{IM}}}}) + {C_2})P G_{\mathrm{eff}}}}{{\sigma _n^2}}} \right)}
\label{eq26}
\end{equation}
\begin{equation}
{E(\rho ,{K_{{\text{TR}}}},{K_{{\text{IM}}}}) = ({C_1}({K_{{\text{TR}}}} - {\beta _1}{K_{{\text{IM}}}}) + {C_2})({C_3}({K_{{\text{IM}}}} - {\beta _2}{K_{{\text{TR}}}}) + {C_4})(1 - \rho )P G_{\mathrm{eff}}}
\label{eq27}
\end{equation}
\end{figure*}

It is emphasized that the above models are locally valid first-order approximations introduced for analytical tractability in the subcarrier-allocation problem, rather than global physical laws. The coupling terms reflect the inherent tradeoff between the two subcarrier groups: increasing $K_{\text{TR}}$ helps reduce transmit-side PAPR and improve PA efficiency, but may suppress receive-side peaks and limit rectifier performance; increasing $K_{\text{IM}}$ enhances receive-side peak formation, but may degrade transmit-side PAPR control. To preserve the validity of the linearization and avoid strongly nonlinear operating regimes (e.g., deep PA saturation or rectifier breakdown), both $K_{\text{TR}}$ and $K_{\text{IM}}$ are restricted to $[0, K/4]$, within which the efficiency variations remain smooth and higher-order terms are negligible. Under this regime, the information-transmission and energy-harvesting performances can be expressed as functions of $(K_{\text{TR}}, K_{\text{IM}})$ and the power-splitting ratio $\rho$, as given in (\ref{eq26}) and (\ref{eq27}).

Consistent with the role of $\Xi$ in (\ref{papr_metric}), the proposed method is implemented as a hierarchically coupled design under the same overall system model. Specifically, $(K_{\text{TR}},K_{\text{IM}})$ determines the system-level tradeoff among receive-side peak enhancement, transmit-side peak-reduction capability, and information transmission, while the TR waveform variable $X_{\mathrm{TR}}$ is optimized subsequently to realize the corresponding transmit-side PAPR suppression under the obtained allocation. Since the power-splitting ratio $\rho$ is typically preset by hardware and service requirements in practical SWIPT receivers, it is treated as fixed in the allocation stage. Under this setting, the resulting allocation problem, subject to a minimum harvested-power requirement, is formulated as

\begin{equation}
\label{eq:P1}
\begin{aligned}
  (\mathrm{P1})\; & \max_{K_{\mathrm{TR}},\,K_{\mathrm{IM}}} \;
  R(K_{\mathrm{TR}},K_{\mathrm{IM}}) \\
  & \text{s.t.} \quad 0 \leq K_{\mathrm{TR}} \leq \tfrac{K}{4}, \\
  & \qquad\;\; 0 \leq K_{\mathrm{IM}} \leq \tfrac{K}{4}, \\
  & \qquad\;\; E(K_{\mathrm{TR}},K_{\mathrm{IM}}) \geq P_{\min}.
\end{aligned}
\end{equation}
where the upper bound $K/4$
follows from the adopted three-group subcarrier partition, which limits the
sizes of the TR and IM subsets to moderate proportions so as to preserve a
sufficient number of QAM data subcarriers.

Since $K_{\mathrm{TR}}$ and $K_{\mathrm{IM}}$ are integer-valued in the
original allocation problem, we next introduce a continuous relaxation for
tractable analysis and optimization. Specifically, define the relaxed
decision vector as
\[
\mathbf{k} = [\,K_{\mathrm{TR}},\; K_{\mathrm{IM}}\,]^T .
\]

With this notation, (P1) can be rewritten in vector form as (P2), where the
objective and the harvested-power constraint are expressed compactly in terms
of affine and quadratic forms of $\mathbf{k}$. This reformulation streamlines
the notation and prepares the subsequent local convex approximation.

Under the continuous relaxation introduced above, and using the local affine
surrogate models for $\eta_{\mathrm{PA}}$ and $\eta_{\mathrm{R}}$, problem
$(\mathrm{P1})$ can be rewritten in the compact vector form
\begin{equation}
\label{eq:P2}
\begin{aligned}
  (\mathrm{P2})\; & \max_{\mathbf{k}} \;
  \Big( N_s K + \mathbf{q}_{0}^{T}\mathbf{k} \Big)
  \log_{2}\!\left( r_0 + \mathbf{q}_{1}^{T}\mathbf{k} \right)  \\
  & \text{s.t.} \quad
  \big(\mathbf{q}_{2}^{T}\mathbf{k} + C_{2}\big)
  \big(\mathbf{q}_{3}^{T}\mathbf{k} + C_{4}\big)
  \geq \eta_{\min},  \\
  & \qquad\;\; \mathbf{0} \preceq \mathbf{k} \preceq \tfrac{K}{4}\mathbf{1},
\end{aligned}
\end{equation}
where $\mathbf{k}=[K_{\mathrm{TR}},\,K_{\mathrm{IM}}]^T$, and
\begin{equation}
\label{eq:q_definitions}
\begin{aligned}
  \eta_{\min}
  &= \frac{P_{\min}}{(1-\rho)P G_{\mathrm{eff}}}, \\
  r_0
  &= 1 + \frac{\rho C_2 P G_{\mathrm{eff}}}{\sigma_n^2}, \\
  \mathbf{q}_0
  &= [-N_s,\; -\mu N_s]^T, \\
  \mathbf{q}_1
  &= \left[
      \frac{\rho C_1 P G_{\mathrm{eff}}}{\sigma_n^2},\;
      -\frac{\rho C_1 \beta_1 P G_{\mathrm{eff}}}{\sigma_n^2}
     \right]^T, \\
  \mathbf{q}_2
  &= [\,C_1,\; -C_1\beta_1\,]^T, \\
  \mathbf{q}_3
  &= [\,-C_3\beta_2,\; C_3\,]^T.
\end{aligned}
\end{equation}

Here, $G_{\mathrm{eff}}$ denotes the effective channel power gain after
precoding/combining and power splitting, introduced to represent the
equivalent scalar channel appearing in the harvested-power and achievable-rate
expressions.

Problem $(\mathrm{P2})$ remains nonconvex due to the coupled product form in
the objective and the nonlinear logarithmic term. To obtain a tractable
iterative solution, we adopt a successive convex approximation (SCA)
procedure. At iteration $t$, given a feasible point $\mathbf{k}^{(t)}$, define
\begin{equation}
\begin{aligned}
s^{(t)} = r_0 + \mathbf{q}_1^T \mathbf{k}^{(t)}.
\end{aligned}
\end{equation}

Since $\log_2(\cdot)$ is concave and continuously differentiable, its
first-order Taylor expansion at $\mathbf{k}^{(t)}$ gives the local affine
approximation
\begin{equation}
\label{eq:log_taylor}
\begin{aligned}
\log_2\!\left(r_0+\mathbf{q}_1^T\mathbf{k}\right)
\approx
\log_2 s^{(t)}
+
\frac{\mathbf{q}_1^T(\mathbf{k}-\mathbf{k}^{(t)})}{s^{(t)}\ln 2}.
\end{aligned}
\end{equation}

This approximation is exact at $\mathbf{k}=\mathbf{k}^{(t)}$ and is used to
construct the convex subproblem solved at the $t$th SCA iteration.

Substituting the local approximation \eqref{eq:log_taylor} into
$(\mathrm{P2})$ yields the following quadratically constrained convex
subproblem at iteration $t$:
\begin{equation}
\label{eq:P3}
\begin{aligned}
  (\mathrm{P3})^{(t)}\; & \max_{\mathbf{k}} \;\;
    \mathbf{k}^{T}\mathbf{P}_{0}^{(t)}\mathbf{k}
    + \tilde{\mathbf{q}}_{0}^{(t)T}\mathbf{k}
    + \tilde{r}_{0}^{(t)} \\
  & \text{s.t.} \quad
    \mathbf{k}^{T}\mathbf{P}_{1}\mathbf{k}
    + \tilde{\mathbf{q}}_{1}^{T}\mathbf{k}
    + \tilde{r}_{1} \geq 0, \\
  & \qquad\;\; \mathbf{0} \preceq \mathbf{k} \preceq \tfrac{K}{4}\mathbf{1}.
\end{aligned}
\end{equation}

The quadratic, linear, and constant coefficients in
$(\mathrm{P3})^{(t)}$ are given by
\begin{equation}
\label{eq:P3_coefficients}
\begin{aligned}
  \mathbf{P}_{0}^{(t)}
  &= \frac{1}{2 s^{(t)} \ln 2}
     \left( \mathbf{q}_{0}\mathbf{q}_{1}^{T}
     + \mathbf{q}_{1}\mathbf{q}_{0}^{T} \right), \\
  \tilde{\mathbf{q}}_{0}^{(t)T}
  &= (\log_{2} s^{(t)})\,\mathbf{q}_{0}^{T}
   + \frac{N_{s}K}{s^{(t)}\ln 2}\,\mathbf{q}_{1}^{T}
   - \frac{\mathbf{q}_{1}^{T}\mathbf{k}^{(t)}}{s^{(t)}\ln 2}\,\mathbf{q}_{0}^{T}, \\
  \tilde{r}_{0}^{(t)}
  &= N_{s}K \log_{2} s^{(t)}
   - \frac{N_{s}K}{s^{(t)}\ln 2}\,\mathbf{q}_{1}^{T}\mathbf{k}^{(t)}, \\
  s^{(t)}
  &= r_0 + \mathbf{q}_1^{T}\mathbf{k}^{(t)}, \\
  \mathbf{P}_{1}
  &= \frac{1}{2}
     \left( \mathbf{q}_{2}\mathbf{q}_{3}^{T}
     + \mathbf{q}_{3}\mathbf{q}_{2}^{T} \right), \\
  \tilde{\mathbf{q}}_{1}^{T}
  &= C_{2}\mathbf{q}_{3}^{T} + C_{4}\mathbf{q}_{2}^{T}, \\
  \tilde{r}_{1}
  &= C_{2}C_{4} - \eta_{\min}.
\end{aligned}
\end{equation}

Here, $\mathbf{P}_{0}^{(t)}$ is obtained by expanding the bilinear term
$\big(\mathbf{q}_{0}^{T}\mathbf{k}\big)
\big(\mathbf{q}_{1}^{T}\mathbf{k}\big)$ into a symmetric quadratic form,
whereas $\mathbf{P}_{1}$ follows from the expansion of
$\big(\mathbf{q}_{2}^{T}\mathbf{k}+C_{2}\big)
\big(\mathbf{q}_{3}^{T}\mathbf{k}+C_{4}\big)$.

Within the considered operating region, $\mathbf{P}_{0}^{(t)} \preceq 0$
and $\mathbf{P}_{1} \preceq 0$. Hence, the objective function in
$(\mathrm{P3})^{(t)}$ is concave quadratic, while the quadratic constraint
defines the superlevel set of a concave function and is therefore convex.
Together with the box constraint, $(\mathrm{P3})^{(t)}$ is a convex
optimization problem. Moreover, whenever a strictly feasible point exists,
Slater's condition is satisfied and strong duality holds.

Accordingly, the optimizer of $(\mathrm{P3})^{(t)}$ can be characterized by
its Karush--Kuhn--Tucker (KKT) conditions. Introducing the Lagrange
multiplier $\lambda \geq 0$ associated with the quadratic constraint, the
stationarity condition is written as
\begin{equation}
2\!\left(\mathbf{P}_{0}^{(t)}+\lambda\mathbf{P}_{1}\right)\mathbf{k}
+\tilde{\mathbf{q}}_{0}^{(t)}+\lambda\tilde{\mathbf{q}}_{1}
-\boldsymbol{\mu}+\boldsymbol{\nu} = \mathbf{0},
\label{eq:kkt_stationarity}
\end{equation}
where $\boldsymbol{\mu}\succeq \mathbf{0}$ and
$\boldsymbol{\nu}\succeq \mathbf{0}$ are the multipliers corresponding to
the lower and upper box constraints, respectively. The complementary-slackness
conditions are
\begin{equation}
\label{eq:kkt_cs}
\begin{aligned}
\mu_i k_i &= 0,\qquad i=1,2, \\
\nu_i\!\left(k_i-\tfrac{K}{4}\right) &= 0,\qquad i=1,2.
\end{aligned}
\end{equation}
together with primal and dual feasibility.

In view of the above KKT system, $(\mathrm{P3})^{(t)}$ can be solved
efficiently by a standard convex optimization routine. Since the decision
vector $\mathbf{k}$ is only two-dimensional, the computational overhead of
this step is negligible in practice. The resulting optimizer is denoted by
$\mathbf{k}^{(t+1)}$ and is used as the new linearization point for the next
SCA iteration.

The above procedure is repeated until
\begin{equation}
\|\mathbf{k}^{(t+1)}-\mathbf{k}^{(t)}\| \le \varepsilon,
\end{equation}
for a prescribed tolerance $\varepsilon$. In all simulated cases considered
in this paper, the proposed iterative procedure converges within a small
number of iterations and yields a stable subcarrier allocation solution.

\begin{algorithm}[t]
\caption{SCA-Based Joint Subcarrier Allocation with TR and IM for MIMO-OFDM SWIPT}
\label{alg:formula_tr_im_eng}
\begin{algorithmic}[1]

\State \textbf{Input:} Effective channel gain $G_{\mathrm{eff}}$ (or channel matrix $\mathbf{H}$); power-splitting ratio $\rho$; minimum harvested power $P_{\min}$; input bits $\mathbf{b}$; IFFT size $K$; QAM order $M$; tolerance $\varepsilon$.
\State \textbf{Output:} Subcarrier allocation $\mathbf{k}^{\star}$; harvested DC power; transmit- and receive-side PAPR.
\Statex

\State \textbf{Initialization:}
Choose a feasible initial point $\mathbf{k}^{(0)}$ satisfying
$\mathbf{0} \preceq \mathbf{k}^{(0)} \preceq \frac{K}{4}\mathbf{1}$ and the harvested-power constraint; set $t=0$.

\Repeat

\State Compute $s^{(t)} = r_0 + \mathbf{q}_1^T \mathbf{k}^{(t)}$.

\State Construct the iteration-dependent coefficients
$\mathbf{P}_0^{(t)}$, $\tilde{\mathbf{q}}_0^{(t)}$, and $\tilde r_0^{(t)}$ according to \eqref{eq:P3_coefficients}.

\State Solve the convex subproblem $(\mathrm{P3})^{(t)}$ using a standard convex optimization routine, and obtain the updated relaxed allocation $\mathbf{k}^{(t+1)}$.

\State Update $t \leftarrow t+1$.

\Until{$\|\mathbf{k}^{(t)}-\mathbf{k}^{(t-1)}\| \le \varepsilon$}

\State Set $\mathbf{k}^{\star} = \mathbf{k}^{(t)}$.

\State Map the relaxed solution $\mathbf{k}^{\star}$ to a feasible integer pair $(K_{\mathrm{TR}},K_{\mathrm{IM}})$, and perform feasibility correction if needed.

\State Partition the subcarriers into $\mathcal{K}_{\mathrm{TR}}$, $\mathcal{K}_{\mathrm{IM}}$, and $\mathcal{K}_{\mathrm{QAM}}$.

\State Perform bit mapping, with QAM symbols on $\mathcal{K}_{\mathrm{QAM}}$ and index modulation on $\mathcal{K}_{\mathrm{IM}}$.

\State Apply tone reservation on $\mathcal{K}_{\mathrm{TR}}$ to reduce the transmit-side PAPR:
\[
\min_{\mathbf{c}_i}\;
\big\|\mathbf{F}^{H}(\mathbf{x}_i+\mathbf{c}_i)\big\|_{\infty}.
\]

\State Generate the time-domain signal by IFFT and pass it through the LDMOS PA model in Fig.~\ref{fig:HPA_combined}.

\State Apply power splitting at the receiver and compute the harvested DC power using the rectifier model in Fig.~\ref{fig:rectifier_combined}.

\State Evaluate the resulting transmit-side and receive-side PAPR.

\end{algorithmic}
\end{algorithm}

As summarized in \textbf{Algorithm~\ref{alg:formula_tr_im_eng}}, the proposed
framework first determines the relaxed subcarrier allocation through an
SCA-based iterative optimization, where each iteration solves the convex
subproblem $(\mathrm{P3})^{(t)}$. The resulting relaxed allocation is then
mapped to an integer-valued subcarrier split, based on which the TR, IM, and
QAM subsets are constructed for waveform generation, PA processing, and
rectifier-side energy evaluation.

\section{Simulation Results}

This section presents simulation results for the proposed SWIPT system incorporating both the nonlinear power amplifier and rectifier models. A path loss of 45~dB is assumed, and regularized zero-forcing (RZF) precoding is employed for MIMO transmission. Antenna and beamforming gains are excluded to focus on fundamental performance. The detailed simulation parameters are summarized in Table~\ref{tab:simulation_parameters}.

\begin{table}[!t]
\centering
\caption{Simulation Parameters}
\begin{tabular}{@{}lc@{}}
\toprule
\multicolumn{2}{c}{Parameter Settings} \\ \midrule
OFDM Subcarriers & 1024 \\
Oversampling Factor & 8 \\
Subcarrier Spacing & 15 kHz \\
Channel Type & TDL-C \\
Number of Transmit Antennas & 4 \\
Number of Receive Antennas & 4 \\
Modulation Order & QPSK \\
PA Input Power (Operating Region) & 12--17 dBm \\
Diode Parameters & \\
\quad Ideality Factor ($\eta$) & 1.05 \\
\quad Thermal Voltage ($V_0$) & 0.026 V \\
\quad Reverse Saturation Current ($I_0$) & $3 \times 10^{-6}$ A \\
\quad Breakdown Saturation Current ($I_{BV}$) & $3 \times 10^{-4}$ A \\
\quad Reverse Breakdown Voltage ($V_B$) & 3.8 V \\ 
\bottomrule
\end{tabular}
\label{tab:simulation_parameters}
\end{table}

The detailed workflow of the transceiver is summarized in \textbf{Algorithm~3}, where subcarrier allocation is determined via KKT optimization, followed by bit mapping, tone reservation, IFFT, and LDMOS amplification. At the receiver, the signal is divided into information decoding and energy harvesting branches, with the rectifier model in Fig.~5 used to compute the harvested DC power, while the PAPR performance is also evaluated.

\subsection{Transmit-Side PAPR Reduction and Power Amplifier Efficiency Improvement}

\begin{figure}[!t]
\centering

\begin{minipage}{0.9\linewidth}
\centering
\includegraphics[width=0.75\linewidth]{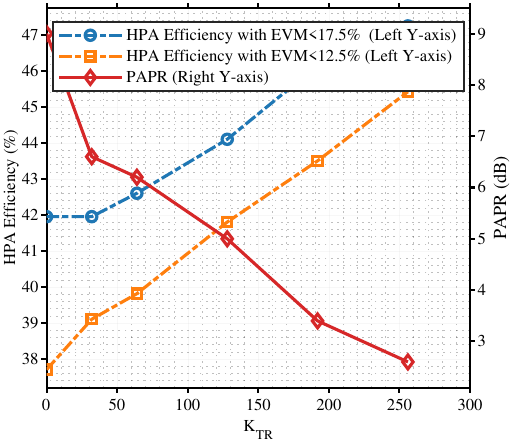}\\
\vspace{3pt}
{\footnotesize\textbf{(a)} Transmit PAPR and PA efficiency versus $K_{\text{TR}}$.}
\end{minipage}

\vspace{8pt}

\begin{minipage}{0.9\linewidth}
\centering
\includegraphics[width=0.75\linewidth]{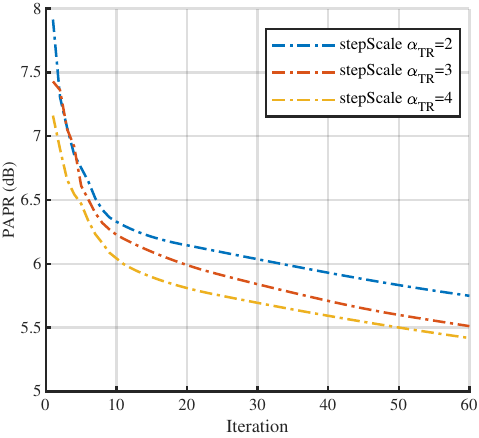}\\
\vspace{3pt}
{\footnotesize\textbf{(b)} Convergence behavior of the TR optimization in Algorithm 2.}
\end{minipage}

\caption{Impact of TR design: (a) transmit PAPR and PA efficiency versus $K_{\text{TR}}$; (b) convergence of the TR optimization algorithm.}
\label{fig8}
\end{figure}

\begin{figure}[!t]
\centering
\includegraphics[width=6cm]{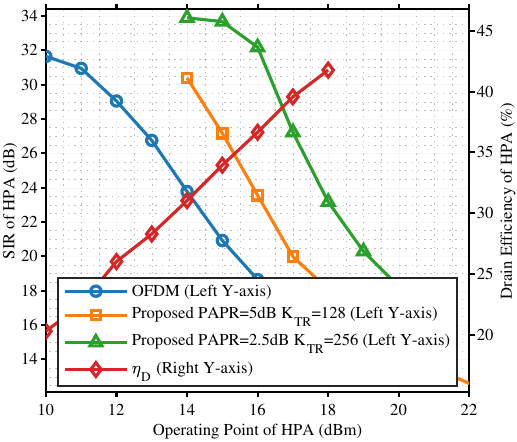}
\caption{SIR Performance under Different PA Operating Points for Proposed TR Scheme and OFDM.}
\label{fig9}
\end{figure}

Fig.~\ref{fig8} illustrates the impact of the TR design. 
Subfigure (a) shows the effect of the number of reserved subcarriers on the transmit PAPR and PA efficiency. 
As the number of reserved subcarriers increases, the transmit PAPR is significantly reduced, 
allowing the PA to operate closer to its saturation region with smaller output back-off. 
When the nonlinear distortion of the PA, measured by the EVM, 
is constrained to practical levels such as 12.5\% or 17.5\%, the amplifier achieves noticeably higher efficiency. Subfigure (b) illustrates the convergence behavior of the proposed TR optimization algorithm under different gradient step sizes. 
Three curves corresponding to different step sizes depict the relationship between the iteration number and the achieved PAPR reduction. 
The PAPR decreases rapidly during the initial iterations and gradually stabilizes as the algorithm converges. 
While larger step sizes accelerate the initial convergence, moderate step sizes provide more stable convergence behavior.

Fig.~\ref{fig9} further evaluates the signal-to-interference ratio (SIR) under different PA operating points for both the proposed scheme (with $K_{\text{TR}}=128$ and $K_{\text{TR}}=256$) and conventional OFDM. The results demonstrate that reducing PAPR not only improves PA efficiency but also alleviates nonlinear distortion, thereby enhancing SIR performance. Together, Fig.~\ref{fig8} and Fig.~\ref{fig9} validate the effectiveness of the proposed method in improving transmit-side energy efficiency by jointly optimizing waveform PAPR and PA operating conditions.

\subsection{Receive-Side PAPR Enhancement and Rectifier Performance}

\begin{figure}[!t]
\centering
\includegraphics[width=6cm]{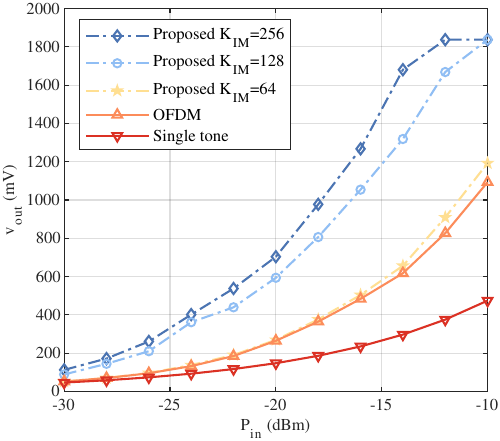}
\caption{Rectifier output voltage comparison under identical RF input power with $K_{\text{TR}}=128$.}
\label{fig10}
\end{figure}

\begin{figure}[!t]
\centering
\includegraphics[width=5.8cm]{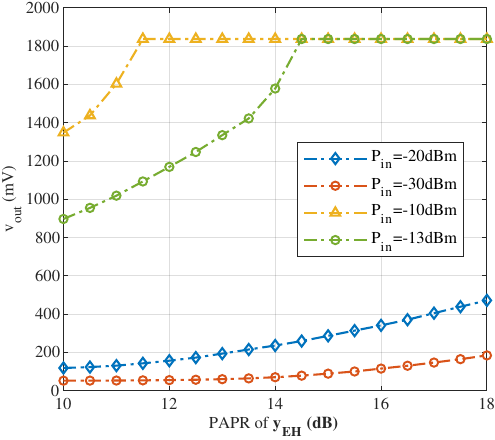}
\caption{Rectifier output voltage versus signal PAPR under different RF input power levels.}
\label{fig11}
\end{figure}

Fig.~\ref{fig10} and Fig.~\ref{fig11} evaluate the rectifier performance based on the nonlinear model in (\ref{eq6})–(\ref{eq7}). In particular, Fig.~\ref{fig10} compares the rectifier output voltage of the proposed scheme with conventional OFDM and single-tone signals under identical RF input power. The proposed waveform achieves a markedly higher output voltage than OFDM. For instance, when the number of index-modulated subcarriers is set to $K_{\text{IM}}=128$ or $256$, the rectifier output voltage is more than twice that of the OFDM baseline.

This improvement is consistent with the theoretical insight of (\ref{eq7}), which establishes a positive correlation between rectifier efficiency and the PAPR of the input signal $\mathbf{y}_{\text{EH}}$. Moreover, the results reveal a fundamental trade-off: by sacrificing only 25\%–35\% of spectral efficiency, the proposed design yields a multiplicative gain in rectification efficiency. This demonstrates that even within the widely adopted MIMO-OFDM communication framework, the proposed waveform design can significantly enhance energy transfer performance, thereby improving the overall efficiency of SWIPT.

\begin{figure}[!t]
\centering
\includegraphics[width=6cm]{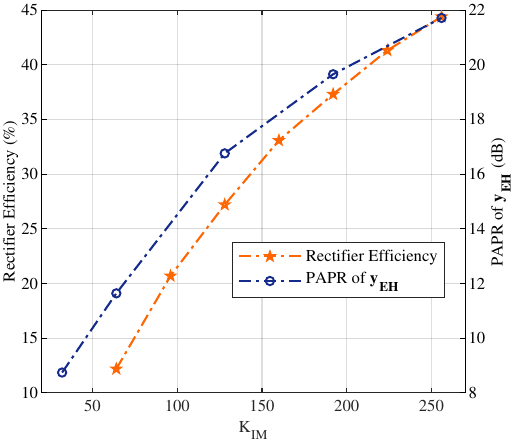}
\caption{PAPR and rectifier performance versus $K_{\text{IM}}$ with $K_{\text{TR}}=128$.}
\label{fig12}
\end{figure}

Fig.~\ref{fig12} further investigates the impact of PAPR on rectifier behavior, validating the analytical relationship derived in (\ref{eq24}). The simulation results demonstrate that, across different RF input power levels, the rectifier output voltage—and consequently the conversion efficiency—increases proportionally with the PAPR of the received waveform $\mathbf{y}_{\text{EH}}$. These observations provide empirical support for the analytical conclusion that high-PAPR signals enhance nonlinear rectifier performance, underscoring the pivotal role of waveform design in enabling efficient wireless power transfer.

\subsection{End-to-End Efficiency Performance for SWIPT Transmission}

\begin{figure}[!t]
\centering
\includegraphics[width=6cm]{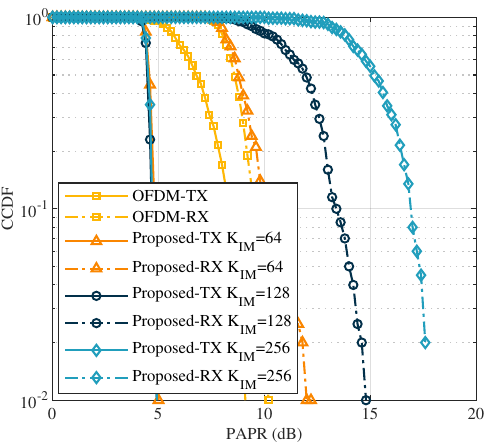}
\caption{Transmit- and receive-side PAPR performance under different subcarrier allocation schemes with $K_{\text{TR}}=128$.}
\label{fig13}
\end{figure}

Fig.~\ref{fig13} illustrates the PAPR performance at both the transmitter and receiver under different subcarrier allocation schemes. On the transmit side, the proposed scheme effectively reduces the PAPR of the OFDM signal to approximately 5~dB, representing a significant improvement over conventional designs. On the receive side, the energy signal obtained through combiner-based merging exhibits a notable increase in PAPR. This enhancement follows the theoretical trend in (\ref{eq24}), confirming that the PAPR gain increases with the number of index-modulated subcarriers. These results verify that the proposed structured subcarrier allocation simultaneously mitigates nonlinear distortion at the transmitter and improves power harvesting capability at the receiver. Having achieved the desired PAPR characteristics, we next evaluate the end-to-end energy transfer performance.

\begin{figure}[!t]
\centering
\includegraphics[width=6cm]{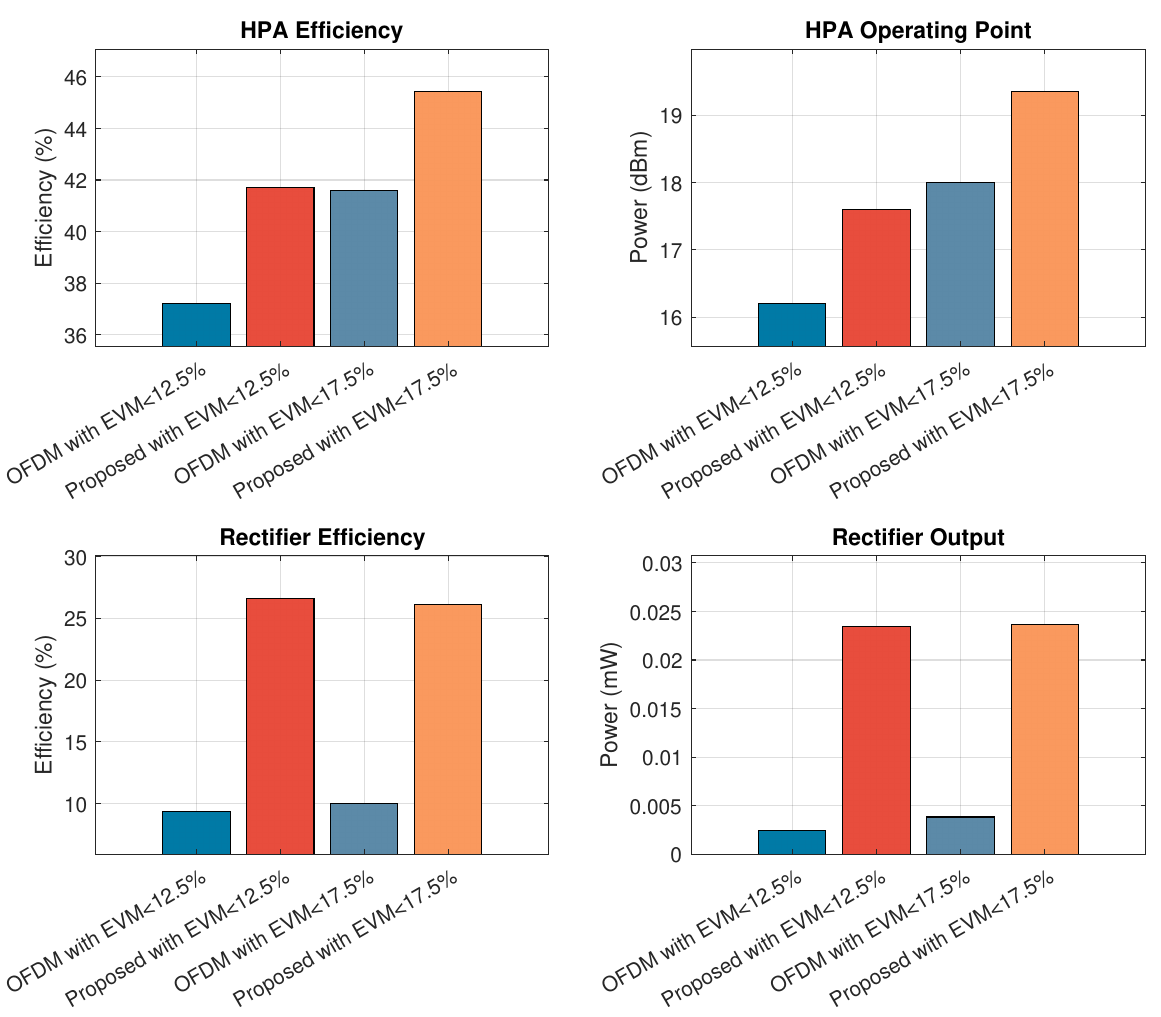}
\caption{End-to-end energy transfer performance comparison between the proposed scheme and OFDM under $K_{\text{TR}}=128$ and $K_{\text{IM}}=128$, with constrained PA operating points based on EVM requirements.}
\label{fig14}
\end{figure}

Fig.~\ref{fig14} presents a comparison of the end-to-end energy transfer performance of the proposed scheme and conventional OFDM under $K_{\text{TR}}=128$ and $K_{\text{IM}}=128$, corresponding to a spectral efficiency reduction of slightly less than 25\%. From an energy-efficiency perspective, it is desirable for the PA to operate close to saturation. However, nonlinear distortion introduces a trade-off between efficiency and signal integrity. To ensure fairness, the PA operation is constrained to meet specific EVM requirements while maximizing the input power. Under these conditions, we evaluate PA operating point, PA efficiency, rectifier efficiency after multipath propagation, and final rectifier output power. The results show that the proposed scheme improves PA efficiency by more than 10\% and nearly triples rectification efficiency at the receiver. Consequently, the overall end-to-end harvested energy is increased by several times, demonstrating the effectiveness of the proposed method in jointly optimizing transmitter- and receiver-side performance.

\begin{figure}[!t]
\centering
\includegraphics[width=6cm]{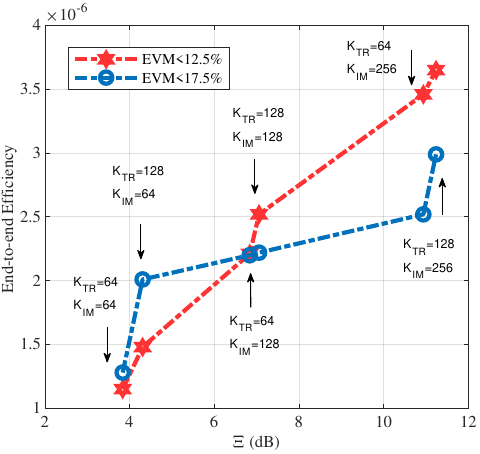}
\caption{Simulated validation of PAPR metric $\Xi$.}
\label{fig15}
\end{figure}

Fig.~\ref{fig15} validates the proposed PAPR-based metric $\Xi$ in (\ref{papr_metric}), defined as the ratio between the average RX-PAPR and TX-PAPR and hence dimensionless. The results show a clear monotonic trend between $\Xi$ and the end-to-end energy efficiency $\eta=\eta_{\mathrm{PA}}\eta_{\mathrm{R}}$ under the considered PA operating range and rectifier model. In particular, larger $\Xi$ is consistently associated with higher overall energy-transfer efficiency, supporting its use as a design-oriented indicator for waveform/resource allocation. The PA input operating region (12--17 dBm) is selected according to the LDMOS characteristics in Fig.~\ref{fig:HPA_combined}, where the working point is set as high as possible while satisfying the EVM constraint to approach the high-efficiency region.

Moreover, in the proposed scheme, $\Xi$ can be flexibly tuned through spectral allocation strategies. This adaptability not only supports efficient energy delivery but also provides an additional degree of freedom for balancing information and energy transmission. In practical SWIPT systems where the power-splitting ratio $\rho$ cannot be dynamically adjusted due to hardware constraints, controlling $\Xi$ through spectral resource allocation offers an effective alternative for realizing energy–information trade-offs.

\begin{figure}[!t]
\centering
\includegraphics[width=6cm]{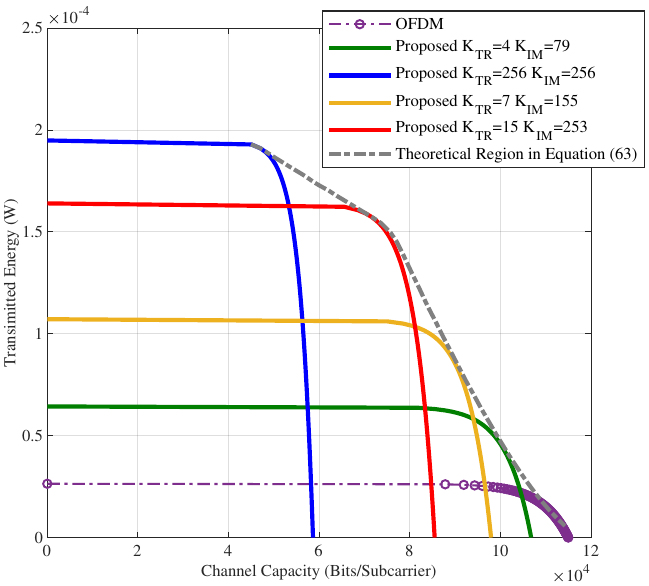}
\caption{Comparison of achievable R–E regions between the proposed scheme and conventional OFDM, along with the theoretical upper bound.}
\label{fig16}
\end{figure}

To quantify the tradeoff between information transmission and energy
harvesting, we define the following normalized scalarized objective:
\begin{equation}
J(\rho,K_{\mathrm{TR}},K_{\mathrm{IM}})
=
\frac{R(\rho,K_{\mathrm{TR}},K_{\mathrm{IM}})}{R_{\max}}
+
\frac{E(\rho,K_{\mathrm{TR}},K_{\mathrm{IM}})}{E_{\max}},
\end{equation}
where, $R_{\max}$ and $E_{\max}$ denote the corresponding normalization
constants, representing the maximum achievable rate and the maximum
harvested-energy value, respectively, within the considered parameter range.

Accordingly, the corresponding R--E tradeoff envelope is characterized by
\begin{equation}
\mathcal{B}_{\mathrm{RE}}
=
\bigcup_{\rho \in [0,1]}
\left\{
\begin{array}{l}
\displaystyle
\max_{K_{\mathrm{TR}},\,K_{\mathrm{IM}}}
J(\rho,K_{\mathrm{TR}},K_{\mathrm{IM}}) \\[6pt]
\text{s.t.}\quad
0 \le K_{\mathrm{TR}},\,K_{\mathrm{IM}} \le \tfrac{K}{4}.
\end{array}
\right\}.
\end{equation}

Fig.~\ref{fig16} depicts the simulated R–E regions of the proposed scheme under different parameter settings, together with the theoretical boundary. Compared with conventional OFDM-based SWIPT, the proposed approach achieves a substantially enlarged R–E region, primarily due to its superior energy efficiency. Furthermore, the results highlight the flexibility of the framework: by tuning the number of TR and IM subcarriers, along with the power-splitting ratio $\rho$, the R–E region can be adaptively shaped to meet diverse system requirements and trade-offs. This adaptability renders the scheme well suited for a wide range of application scenarios. Finally, the close agreement between the theoretical upper bound and the simulated results validates both the accuracy of the analytical model and the practical effectiveness of the proposed joint optimization strategy.

\section{Conclusion}

This paper has investigated the pivotal role of PAPR in SWIPT systems, with particular emphasis on its dual impact on power amplifier efficiency and rectifier performance. By developing a unified analytical framework, it has been shown that the end-to-end energy efficiency is fundamentally governed by the PAPR characteristics of the transmit waveform, thereby revealing an inherent trade-off between reliable information transfer and efficient energy harvesting.  

To overcome this challenge, a novel SWIPT architecture with flexible frequency-domain resource allocation was proposed, where the available spectrum is partitioned into PAPR-controlled, peak-shaping, and data-transmission blocks. This design enables adaptive balancing between amplifier efficiency and rectification effectiveness. Extensive simulations have validated the proposed approach, demonstrating substantial gains in harvested energy without compromising data transmission reliability.  

In summary, the proposed framework provides new insights into waveform design and resource allocation for SWIPT, highlighting the necessity of incorporating nonlinear device characteristics into end-to-end system optimization. Future research will focus on extending the framework to multi-user and large-scale MIMO scenarios, as well as investigating machine learning-based strategies for real-time adaptive optimization of energy–information trade-offs.

\bibliographystyle{IEEEtran}
\bibliography{SWIPT_F}

\begin{thebibliography}{10}
\providecommand{\url}[1]{#1}
\csname url@samestyle\endcsname
\providecommand{\newblock}{\relax}
\providecommand{\bibinfo}[2]{#2}
\providecommand{\BIBentrySTDinterwordspacing}{\spaceskip=0pt\relax}
\providecommand{\BIBentryALTinterwordstretchfactor}{4}
\providecommand{\BIBentryALTinterwordspacing}{\spaceskip=\fontdimen2\font plus
\BIBentryALTinterwordstretchfactor\fontdimen3\font minus \fontdimen4\font\relax}
\providecommand{\BIBforeignlanguage}[2]{{%
\expandafter\ifx\csname l@#1\endcsname\relax
\typeout{** WARNING: IEEEtran.bst: No hyphenation pattern has been}%
\typeout{** loaded for the language `#1'. Using the pattern for}%
\typeout{** the default language instead.}%
\else
\language=\csname l@#1\endcsname
\fi
#2}}
\providecommand{\BIBdecl}{\relax}
\BIBdecl

\bibitem{choi2020simultaneous}
K.~W. Choi, S.~I. Hwang, A.~A. Aziz, H.~H. Jang, J.~S. Kim, D.~S. Kang, and D.~I. Kim, ``Simultaneous wireless information and power transfer ({SWIPT}) for internet of things: Novel receiver design and experimental validation,'' \emph{IEEE Internet Things J.}, vol.~7, no.~4, pp. 2996--3012, Apr. 2020.

\bibitem{krikidis2014swipt}
I.~Krikidis, S.~Timotheou, S.~Nikolaou, G.~Zheng, D.~W.~K. Ng, and R.~Schober, ``Simultaneous wireless information and power transfer in modern communication systems,'' \emph{IEEE Commun. Mag.}, vol.~52, no.~11, pp. 104--110, 2014.

\bibitem{perera2018survey}
T.~D.~P. Perera, D.~N.~K. Jayakody, S.~K. Sharma, S.~Chatzinotas, and J.~Li, ``Simultaneous wireless information and power transfer {(SWIPT)}: Recent advances and future challenges,'' \emph{IEEE Commun. Surveys Tuts.}, vol.~20, no.~1, pp. 264--302, 2018.

\bibitem{hossain2019survey}
M.~A. Hossain, R.~Md~Noor, K.-L.~A. Yau, I.~Ahmedy, and S.~S. Anjum, ``A survey on simultaneous wireless information and power transfer with cooperative relay and future challenges,'' \emph{IEEE Access}, vol.~7, pp. 19\,166--19\,198, 2019.

\bibitem{8636993}
J.~Tang, J.~Luo, M.~Liu, D.~K.~C. So, E.~Alsusa, G.~Chen, K.~K. Wong, and J.~A. Chambers, ``Energy efficiency optimization for {NOMA} with {SWIPT},'' \emph{IEEE J. Sel. Topics Signal Process.}, vol.~13, no.~3, pp. 452--466, 2019.

\bibitem{6774838}
M.~R.~A. Khandaker and K.~K. Wong, ``{SWIPT} in {MISO} multicasting systems,'' \emph{IEEE Wireless Commun. Lett.}, vol.~3, no.~3, pp. 277--280, 2014.

\bibitem{clerckx2018fundamentals}
B.~Clerckx, R.~Zhang, R.~Schober, D.~W.~K. Ng, D.~I. Kim, and H.~V. Poor, ``Fundamentals of wireless information and power transfer: From {RF} energy harvester models to signal and system designs,'' \emph{IEEE J. Sel. Areas Commun.}, vol.~37, no.~1, pp. 4--33, Sep. 2018.

\bibitem{clerckx2018wireless}
B.~Clerckx, ``Wireless information and power transfer: Nonlinearity, waveform design, and rate-energy tradeoff,'' \emph{IEEE Trans. Signal Process.}, vol.~66, no.~4, pp. 847--862, Feb. 2018.

\bibitem{zhang2013mimo}
R.~Zhang and C.~K. Ho, ``Mimo broadcasting for simultaneous wireless information and power transfer,'' \emph{IEEE Trans. Wireless Commun.}, vol.~12, no.~5, pp. 1989--2001, May 2013.

\bibitem{kim2016new}
D.~I. Kim, J.~H. Moon, and J.~J. Park, ``New {SWIPT} using {PAPR}: How it works,'' \emph{IEEE Wireless Commun. Lett.}, vol.~5, no.~6, pp. 672--675, Dec. 2016.

\bibitem{jang2020novel}
H.~H. Jang, K.~W. Choi, and D.~I. Kim, ``Novel frequency-splitting {SWIPT} for overcoming amplifier nonlinearity,'' \emph{IEEE Wireless Commun. Lett.}, vol.~9, no.~6, pp. 826--829, Jun. 2020.

\bibitem{xu2014multiuser}
J.~Xu, L.~Liu, and R.~Zhang, ``Multiuser miso beamforming for simultaneous wireless information and power transfer,'' \emph{IEEE Trans. Signal Process.}, vol.~62, no.~18, pp. 4798--4810, 2014.

\bibitem{valenta2015theoretical}
C.~R. Valenta, M.~M. Morys, and G.~D. Durgin, ``Theoretical energy-conversion efficiency for energy-harvesting circuits under power-optimized waveform excitation,'' \emph{IEEE Trans. Microw. Theory Techn.}, vol.~63, no.~5, pp. 1758--1767, May 2015.

\bibitem{eidaks2022fast}
J.~Eidaks, R.~Kusnins, R.~Babajans, D.~Cirjulina, J.~Semenjako, and A.~Litvinenko, ``Fast and accurate approach to {RF-DC} conversion efficiency estimation for multi-tone signals,'' \emph{Sensors}, vol.~22, no.~3, p. 787, 2022.

\bibitem{clerckx2016waveform}
B.~Clerckx and E.~Bayguzina, ``Waveform design for wireless power transfer,'' \emph{IEEE Trans. Signal Process.}, vol.~64, no.~23, pp. 6313--6328, Dec. 2016.

\bibitem{abeywickrama2021refined}
S.~Abeywickrama, R.~Zhang, and C.~Yuen, ``Refined nonlinear rectenna modeling and optimal waveform design for multi-user multi-antenna wireless power transfer,'' \emph{IEEE J. Sel. Top. Signal Process.}, vol.~15, no.~5, pp. 1198--1210, Aug. 2021.

\bibitem{zhang2023waveform}
Y.~Zhang and B.~Clerckx, ``Waveform design for wireless power transfer with power amplifier and energy harvester nonlinearities,'' \emph{IEEE Trans. Signal Process.}, vol.~71, pp. 2638--2653, 2023.

\bibitem{andrews2014will}
J.~G. Andrews, S.~Buzzi, W.~Choi, S.~V. Hanly, A.~Lozano, A.~C.~K. Soong, and J.~C. Zhang, ``What will {5G} be?'' \emph{IEEE J. Sel. Areas Commun.}, vol.~32, no.~6, pp. 1065--1082, 2014.

\bibitem{parkvall2018nr}
S.~Parkvall, E.~Dahlman, A.~Furuskar, and M.~Frenne, ``{NR}: The new {5G} radio access technology,'' \emph{IEEE Commun. Standards Mag.}, vol.~1, no.~4, pp. 24--30, 2018.

\bibitem{han2005overview}
S.~H. Han and J.~H. Lee, ``An overview of peak-to-average power ratio reduction techniques for multicarrier transmission,'' \emph{IEEE Wireless Commun.}, vol.~12, no.~2, pp. 56--65, 2005.

\bibitem{jiang2008overview}
T.~Jiang and Y.~Wu, ``An overview: Peak-to-average power ratio reduction techniques for {OFDM} signals,'' \emph{IEEE Trans. Broadcast.}, vol.~54, no.~2, pp. 257--268, 2008.

\bibitem{raab2002power}
F.~H. Raab, P.~Asbeck, S.~Cripps, P.~B. Kenington, Z.~B. Popovic, N.~Pothecary, J.~F. Sevic, and N.~O. Sokal, ``Power amplifiers and transmitters for {RF} and microwave,'' \emph{IEEE Trans. Microw. Theory Techn.}, vol.~50, no.~3, pp. 814--826, 2002.

\bibitem{camarchia2015doherty}
V.~Camarchia, M.~Pirola, R.~Quaglia, S.~Jee, Y.~Cho, and B.~Kim, ``The doherty power amplifier: Review of recent solutions and trends,'' \emph{IEEE Trans. Microw. Theory Techn.}, vol.~63, no.~2, pp. 559--571, 2015.

\bibitem{collado2014optimal}
A.~Collado and A.~Georgiadis, ``Optimal waveforms for efficient wireless power transmission,'' \emph{IEEE Microw. Wireless Compon. Lett.}, vol.~24, no.~5, pp. 354--356, May 2014.

\bibitem{ayir2023}
N.~Ayir, T.~Riihonen, and M.~Heino, ``Practical waveform-to-energy harvesting model and transmit waveform optimization for {RF} wireless power transfer systems,'' \emph{IEEE Trans. Microw. Theory Techn.}, vol.~71, no.~12, pp. 5498--5514, 2023.

\bibitem{litvinenko2018usage}
A.~Litvinenko, J.~Eidaks, and A.~Aboltins, ``Usage of signals with a high {PAPR} level for efficient wireless power transfer,'' in \emph{Proc. IEEE 6th Workshop Adv. Inf., Electron. Electr. Eng. (AIEEE)}, 2018, pp. 1--5.

\bibitem{ayir2021joint}
N.~Ayir and T.~Riihonen, ``Joint impact of input power, {PAPR}, and load resistance on the receiver efficiency of multisine waveforms in {RF} energy harvesting,'' in \emph{Proc. IEEE Wireless Power Transfer Conf. (WPTC)}, 2021, pp. 1--4.

\bibitem{li2011improved}
H.~Li, T.~Jiang, and Y.~Zhou, ``An improved tone reservation scheme with fast convergence for {PAPR} reduction in {OFDM} systems,'' \emph{IEEE Trans. Broadcast.}, vol.~57, no.~4, pp. 902--906, 2011.

\bibitem{jiang2015novel}
T.~Jiang, C.~Ni, C.~Ye, Y.~Wu, and K.~Luo, ``A novel multi-block tone reservation scheme for {PAPR} reduction in {OQAM-OFDM} systems,'' \emph{IEEE Trans. Broadcast.}, vol.~61, no.~4, pp. 717--722, 2015.

\bibitem{wang2019scr}
J.~Wang, X.~Lv, and W.~Wu, ``{SCR}-based tone reservation schemes with fast convergence for {PAPR} reduction in {OFDM} system,'' \emph{IEEE Wireless Commun. Lett.}, vol.~8, no.~2, pp. 624--627, 2019.

\bibitem{li2018tone}
B.~Li, L.~Hu, F.~Yang, L.~Ding, and T.~Song, ``Tone reservation ratio optimization for {PAPR} reduction in {OFDM} systems,'' in \emph{Proc. IEEE Wireless Commun. Netw. Conf. (WCNC)}, 2018, pp. 1--6.

\bibitem{krongold2004active}
B.~S. Krongold and D.~L. Jones, ``An active-set approach for {OFDM} {PAR} reduction via tone reservation,'' \emph{IEEE Trans. Signal Process.}, vol.~52, no.~2, pp. 495--509, 2004.

\bibitem{claessens2019multitone}
S.~Claessens, N.~Pan, D.~Schreurs, and S.~Pollin, ``Multitone {FSK} modulation for {SWIPT},'' \emph{IEEE Trans. Microw. Theory Techn.}, vol.~67, no.~5, pp. 1665--1674, May 2019.

\bibitem{krikidis2020information}
I.~Krikidis, ``Information-energy capacity region for {SWIPT} systems with power amplifier nonlinearity,'' in \emph{Proc. IEEE Int. Symp. Inf. Theory (ISIT)}, 2020, pp. 3067--3072.

\bibitem{ng2013wireless}
D.~W.~K. Ng, E.~S. Lo, and R.~Schober, ``Wireless information and power transfer: Energy efficiency optimization in {OFDMA} systems,'' \emph{IEEE Trans. Wireless Commun.}, vol.~12, no.~12, pp. 6352--6370, 2013.

\bibitem{zhou2014wireless}
X.~Zhou, R.~Zhang, and C.~K. Ho, ``Wireless information and power transfer in multiuser {OFDM} systems,'' \emph{IEEE Trans. Wireless Commun.}, vol.~13, no.~4, pp. 2282--2294, 2014.

\bibitem{krikidis2019tone}
I.~Krikidis and C.~Psomas, ``Tone-index multisine modulation for {SWIPT},'' \emph{IEEE Signal Process. Lett.}, vol.~26, no.~8, pp. 1252--1256, Aug. 2019.

\bibitem{ng2013ofdm}
D.~W.~K. Ng, E.~S. Lo, and R.~Schober, ``Energy-efficient power allocation in {OFDM} systems with wireless information and power transfer,'' in \emph{Proc. IEEE Int. Conf. Commun. (ICC)}, 2013, pp. 4125--4130.

\bibitem{kim2020signal}
J.~Kim, B.~Clerckx, and P.~D. Mitcheson, ``Signal and system design for wireless power transfer: Prototype, experiment and validation,'' \emph{IEEE Trans. Wireless Commun.}, vol.~19, no.~11, pp. 7453--7469, Nov. 2020.

\bibitem{singh2021review}
S.~Singh and J.~Malik, ``Review of efficiency enhancement techniques and linearization techniques for power amplifier,'' \emph{Int. J. Circuit Theory Appl.}, vol.~49, no.~3, pp. 762--777, 2021.

\bibitem{rotenberg2020efficient}
S.~A. Rotenberg, S.~K. Podilchak, P.~D.~H. Re, C.~Mateo-Segura, G.~Goussetis, and J.~Lee, ``Efficient rectifier for wireless power transmission systems,'' \emph{IEEE Trans. Microw. Theory Techn.}, vol.~68, no.~5, pp. 1921--1932, 2020.

\bibitem{hsms285x}
{Avago Technologies}, ``Hsms-285x series surface mount microwave schottky detector diodes datasheet,'' 2007, available: \url{https://www.broadcom.com/}.

\bibitem{zhou2013wireless}
X.~Zhou, R.~Zhang, and C.~K. Ho, ``Wireless information and power transfer: Architecture design and rate-energy tradeoff,'' \emph{IEEE Trans. Commun.}, vol.~61, no.~11, pp. 4754--4767, 2013.

\end{thebibliography}
 
\vspace{11pt}

\vfill

\end{document}